\documentclass[twocolumn,aps,pra,showpacs,raggedbottom,nobalancelastpage,amssymb,superscriptaddress]{revtex4-1}

\usepackage{graphicx}
\usepackage{amsmath}
\usepackage{amssymb}
\usepackage{bm}
\usepackage[usenames]{color}
\usepackage{amsfonts}
\usepackage{appendix}
\usepackage{dsfont}

\newcommand{\ket}[1]{|#1\rangle}
\newcommand{\vt}[1]{\bs{#1}}

\definecolor{Nathanblue}{rgb}{0.,0.24,0.51}

\def\be{\begin{equation}}
\def\ee{\end{equation}}

\def\bs#1{\boldsymbol{#1}}

\begin{document}

\title{ Measurement of Chern numbers through center-of-mass responses}

\author{H. M. Price}
\email[]{hannah.price@unitn.it}
\affiliation{INO-CNR BEC Center and Dipartimento di Fisica, Universit\`{a} di Trento, I-38123 Povo, Italy}
\author{O. Zilberberg}
\affiliation{Institute for Theoretical Physics, ETH Zurich, 8093 Z{\"u}rich, Switzerland}
\author{T. Ozawa}
\affiliation{INO-CNR BEC Center and Dipartimento di Fisica, Universit\`{a} di Trento, I-38123 Povo, Italy}
\author{I. Carusotto}
\affiliation{INO-CNR BEC Center and Dipartimento di Fisica, Universit\`{a} di Trento, I-38123 Povo, Italy}
\author{N. Goldman}
\email[]{ngoldman@ulb.ac.be}
\affiliation{CENOLI, Facult{\'e} des Sciences, Universit{\'e} Libre de Bruxelles (U.L.B.), B-1050 Brussels, Belgium}

\begin{abstract}
 
Probing the center-of-mass of an ultracold atomic cloud can be used to measure Chern numbers, the topological invariants underlying the quantum Hall effects. In this work, we show how such center-of-mass observables can have a much richer dependence on topological invariants than previously discussed. In fact, the response of the center of mass depends not only on the current density, typically measured in a solid-state system, but also on the particle density, which itself can be sensitive to the topology of the band structure. We apply a semiclassical approach, supported by numerical simulations, to highlight the key differences between center-of-mass responses and more standard conductivity measurements. We illustrate this by analyzing both the two- and the four-dimensional quantum Hall effects. These results have important implications for experiments in engineered topological systems, such as ultracold gases and photonics. 
\end{abstract} 

\pacs{67.85.-d, 03.65.Sq, 37.10.Jk, 73.43.-f}

\maketitle

\section{Introduction}

Over recent decades, there has been great interest in studying topological phases of matter~\cite{RMP_TI, RMP_TI2}. In these systems, energy bands can be characterised by topological invariants, which have direct physical consequences in quantised bulk responses and robust edge physics. In the famous 2D quantum Hall (QH) effect, for example, the Hall conductance is quantised in multiples of the integer first Chern number (1CN), an important topological invariant of 2D energy bands~\cite{TKNN}. 

Although the 2D QH effect was first studied for electrons in solid-state materials subject to a perpendicular magnetic field, there has been much progress also in exploring this physics for other analogue systems, such as ultracold atomic gases~\cite{aidelsburger2013,miyake2013,jotzu2014,Aidelsburger:2015, Mancini:2015,Stuhl:2015} and photonics~\cite{Rechtsman, Hafezi}. In these systems, the particles are uncharged, and the effects of the magnetic field must be engineered artificially using other means~\cite{Dalibard2011,Goldman:2014bv, Hafezi_review}. These engineered platforms offer a variety of new opportunities, such as the recent proposal for engineering the 4D QH effect for the first time~\cite{4Datoms:2015, 4Dphotons:2015}. This may be achieved by combining a three-dimensional system of atoms or photons with a ``synthetic" dimension, where internal degrees of freedom are coupled to simulate an additional spatial dimension~\cite{Boada2012, Celi:2014, Mancini:2015,Stuhl:2015}. 

In atomic or photonic systems, different physical concepts are required, not only to access quantum Hall physics, but also to measure and probe its signatures. Usually in a solid-state material, the QH effect is observed in the electrical conductivity, namely, through voltage or current measurements. However, such measurements are difficult to perform in analogue quantum Hall systems, and instead much work has gone into finding new tools for probing the topological and geometrical properties of energy bands~\cite{onur:2008, alba:2011, Price:2012,Dauphin:2013, Price:2013, Liu:2013, atala:2013, abanin:2013, atala:2014, Ozawa:2014, Hafezi:2014, Hauke:2014,Deng:2014, Oded:2014, li:2015, grusdt:2015, 2DSOC2:2015,Duca,Flaschner}. In particular, QH responses can be measured in the center-of-mass-drift of an ultracold atomic cloud~\cite{Price:2012,Dauphin:2013,Aidelsburger:2015, 4Datoms:2015}, or in the displacement of the center-of-mass of the photon steady-state~\cite{4Dphotons:2015, Ozawa:2014}. 

Measurements of observables related to the center-of-mass (c.m.) motion typically depend not just on the current, which is usually measured in a solid-state system, but also on the particle density. For example, the c.m. velocity for a cloud of $N_{\text{tot}}$ atoms with total velocity ${\bs v}_{\text{tot}}$ is defined as
\begin{eqnarray}
{\bs v}_\text{c.m.} = \frac{{\bs v}_{\text{tot}}}{N_{\text{tot}}} = \frac{{\bs j}}{n} ,\label{eq:atomcom}
\end{eqnarray}
where the current density is ${\bs j}={\bs v}_{\text{tot}}/L^{d}$ and the particle density is $n = N_{\text{tot}}/L^{d}$ for a system of length $L$ and dimension $d$. For a QH system, the current density is proportional to the topological invariants (Chern numbers) of filled energy bands [see e.g. Eq.~\eqref{eq:jx2dintro} below]. This behaviour is captured directly by c.m. observables for the cases previously studied~\cite{Price:2012,Dauphin:2013,Aidelsburger:2015,4Dphotons:2015, Ozawa:2014}, in which the particle density $n$ only contributes an unimportant constant factor. However, the particle density can itself become a function of the geometrical and topological properties of the energy bands in the presence of external (possibly artificial) magnetic perturbations~\cite{Xiao2010, Xiao_ED, 4Datoms:2015, onur:2008, Duval}. As we show in this paper, center-of-mass responses can therefore exhibit a much richer dependence on topological invariants than previously discussed. 
 
In this work, we use a semiclassical approach to explore how c.m. responses depend on the topological invariants of filled energy bands. We illustrate these results for the 2D quantum Hall effect, where we show that center-of-mass observables can exhibit nonlinear topological responses, which would not be observed in conductivity measurements, as routinely performed in solid-state systems. Our findings are directly relevant for current experiments in ultracold gases~\cite{aidelsburger2013,miyake2013,jotzu2014,Aidelsburger:2015, Mancini:2015,Stuhl:2015}, photonics~\cite{Rechtsman, Hafezi} and even classical mechanical systems~\cite{Huber}. As far as ultracold gases are concerned, we emphasise that our semiclassical treatment is valid for uniformly-filled bands of either bosons or fermions, as can be achieved through fermionic statistics (for Fermi gases) or through thermal effects. We also build on our recent experimental proposals for the 4D quantum Hall effect~\cite{4Datoms:2015, 4Dphotons:2015} and show that in such 4D systems, our results have important implications for experimental design and detection. We demonstrate, for example, pathological cases in which 4D topological invariants would vanish from center-of-mass responses, while persisting in conductivity measurements. 

\subsubsection*{Main implications for the 2D quantum Hall effect}

Before proceeding, we emphasise the key implications of our results for the 2D quantum Hall effect. For a filled band of electrons, the transverse (Hall) current density follows the linear relation~\cite{TKNN}
\begin{eqnarray}
j^{x} = \frac{e^2}{h} E_y \nu_1^{xy} ,\label{eq:jx2dintro}
\end{eqnarray} 
where $E_y$ is an external electric field aligned along the $y$ direction, $e$ is the elementary charge, $h$ is Planck's constant, and $ \nu_1^{xy} $ is the topological 1CN  of the populated band (defined in Eq.~\eqref{eq:first_chern} below);  in the 2D QH effect, a non-zero 1CN, and hence a non-zero Hall current, is due to an applied perpendicular magnetic field. As the 1CN is topological, the Hall response in Eq.~\eqref{eq:jx2dintro} is remarkably robust; for example, it is insensitive to small changes in the magnetic field provided that the energy gap of the system remains open. This leads to characteristic plateaus in the Hall conductivity $\sigma_{xy} = j^{x}/ E_y$ plotted as a function of the applied magnetic field, where the height of a given plateau is proportional to the 1CN (summed over filled energy bands)~\cite{TKNN}. 

In contrast, c.m. observables can have a much more complicated dependence on the topological index of an energy band. We find that the c.m. velocity \eqref{eq:atomcom} of an ultracold cloud, for example, can include non-linear topological responses, such as a contribution proportional to $(\nu_1^{xy})^2$ under a small perturbation of the applied (artificial) magnetic field [see Eq.~\eqref{eq:taylor_result_2D} below]. Such terms, which have no analogue in current responses, stem from an interplay between the topological invariants appearing in both the current density and particle density in Eq.~(\ref{eq:atomcom}). These effects could be observed in state-of-the-art experiments with ultracold atoms and in photonics~\cite{aidelsburger2013,miyake2013,jotzu2014,Aidelsburger:2015, Mancini:2015,Stuhl:2015,Rechtsman,Hafezi}, where magnetic perturbations can both arise naturally from experimental uncertainties in the (artificial) magnetic flux imposed, and be engineered deliberately using current experimental techniques. Our theory allows one to clearly identify these topological effects and to understand the important differences between probing QH physics through c.m. observables versus conductivity measurements. 

\subsection*{Outline}

The structure of this paper is as follows: we begin by reviewing in Section~\ref{sec:semiclassical} how the semiclassical equations of motion can be used to calculate the quantum Hall current response and other relevant observables. In Section~\ref{sec:2D} we study a 2D quantum Hall system, emphasising the effects of a magnetic field perturbation on the particle density and hence on any c.m. observables. In particular, we illustrate the effects of non-linear topological responses in the c.m. transverse velocity. In Section~\ref{sec:4D}, as a further example, we consider a 4D quantum Hall system, where the distinction between c.m. observables and current measurements has very striking implications for the design of future experiments; for example, we identify pathological configurations of the perturbing electromagnetic fields for which 4D topological invariants can be extracted from a current response, and yet not from a center-of-mass response. Finally, we present additional experimental remarks in Section~\ref{sec:experimental} and draw conclusions in Section~\ref{sec:conclusions}.

\section{The Semiclassical Approach to the Quantum Hall Effect} \label{sec:semiclassical}

In this Section, we develop a semiclassical description of center-of mass responses. In Section~\ref{sec:chern}, we introduce the geometrical and topological properties of eigenstates in an energy band of a quantum Hall system, before reviewing the semiclassical equations of motion for a wave packet constructed out of these eigenstates in Section~\ref{sec:semiclass}~\cite{chang1, chang2, XiaoPol, Gao, Gao2}. From these equations, we review how to derive the modified density of states, the quantum Hall current density and relevant center-of-mass observables in Sections~\ref{sec:modified}, \ref{sec:current} \& \ref{sec:c.m.} respectively. By keeping terms up to second-order in the external perturbing fields, this semiclassical framework can be used to describe both current experiments on 2D quantum Hall physics in ultracold gases and photonics~\cite{aidelsburger2013,miyake2013,jotzu2014,Aidelsburger:2015, Mancini:2015,Stuhl:2015,Rechtsman, Hafezi}, see Section~\ref{sec:2D}, as well as proposed schemes for realising the 4D quantum Hall effect~\cite{4Datoms:2015, 4Dphotons:2015}, see Section~\ref{sec:4D}. 

\subsection{ The Berry Curvature, the First Chern Number and the Second Chern Number} \label{sec:chern}

We begin from a particle in a periodic potential, where the eigenstates can be expressed through Bloch's theorem as $\ket{\chi_{n,\vt{k}}} \!=\! e^{i\vt{k}\cdot \vt{r}} | u_{n,\vt{k}} \rangle$, where $|u_{n,\vt{k}} \rangle$ are the periodic Bloch functions and $\vt{k}$ is the corresponding quasi-momentum. In the Brillouin zone (BZ), the Bloch functions $|u_{n,\vt{k}} \rangle$ form bands where the energy dispersion $\mathcal{E}_n({\bs k})$ is labelled by the band index $n$. In this paper, we focus on the physics of an energetically-isolated non-degenerate energy band, and so hereafter we drop the label $n$. 

The eigenstates that make up the energy band can have nontrivial local geometrical properties as encoded, for example, in the Berry curvature~\cite{Xiao2010}. In this paper, we deal with systems of various dimensionalities and so we express the Berry curvature as a general differential 2-form
\begin{eqnarray}
\Omega &=&\frac{1}{2} \Omega^{\mu \nu}(\bs k) \text{d}k_{\mu}\! \wedge \! \text{d}k_{\nu}, \nonumber \\
\Omega^{\mu \nu} (\bs k)&=&  \partial_{k_\mu} \mathcal{A}_{k_\nu }  -  \partial_{k_\nu} \mathcal{A}_{k_\mu}, \label{eq:berry2form}
\end{eqnarray}
where $\wedge$ is the antisymmetric wedge product, $\mathcal{A}_{k_\mu} = i \langle u_{\vt{k}} |\partial_{k_\mu }| u_{\vt{k}} \rangle $ is the Berry connection and where the indices $\mu, \nu$ run over all spatial coordinates with Einstein summation convention. As can be seen from this definition, the Berry curvature components are antisymmetric under an exchange of indices $\Omega^{\mu \nu} (\bs k) = - \Omega^{ \nu \mu} (\bs k)$. Importantly, geometrical properties, such as the Berry curvature, are also closely related to key topological properties of the energy bands. 

In general, topological phases of matter can be classified according to the symmetries and dimensionality of the system~\cite{Ryu}. In this paper, we focus on non-interacting systems without any symmetries, where the energy bands are topologically trivial in odd dimensions but can be characterised by non-zero topological integers in even dimensions~\cite{TKNN, Nakahara, Ryu}. We consider, in particular, systems with two and four dimensions, where the relevant topological invariants are known as the first and second Chern numbers respectively. For a 2D system in, e.g. the $x\!-\!y$ plane, the 1CN is calculated from the Berry curvature as
\begin{align}
\nu_1^{xy}\!&=\!\frac{1}{2 \pi} \int_{\mathbb{T}^2}  \Omega =\!\frac{1}{2 \pi} \int_{\mathbb{T}^2} \Omega^{xy} \text{d}k_{x} \text{d}k_{y}
  \, \, \in \mathbb{Z},   \label{eq:first_chern}
\end{align}
where the integral is over the first (magnetic) two-dimensional BZ, which is denoted here by $\mathbb{T}^2$ to  emphasise that it is topologically equivalent to a 2-torus. We note that $\nu_1^{xy}= - \nu_1^{yx}$ by the antisymmetry of the Berry curvature. 

In 4D, the second Chern number (2CN) can also be calculated from the Berry curvature as~\cite{Avron1988,IQHE4D,QiZhang, Fukui:2008, lian:2016}
\begin{align}
\nu_2\!&=\!\frac{1}{8 \pi^2} \int_{\mathbb{T}^4}  \Omega \wedge \Omega \, \, \in \mathbb{Z}, \notag \\
&=\frac{1}{32 \pi^2} \int_{\mathbb{T}^4}  \epsilon_{\alpha \beta \gamma \delta }  \Omega^{\alpha \beta } \Omega^{\gamma \delta} \text{d}^4k, \notag\\
&=\!\frac{1}{4 \pi^2} \int_{\mathbb{T}^4}  \Omega^{xy}\Omega^{zw}\!+\! \Omega^{wx}\Omega^{zy}\!+\! \Omega^{zx}\Omega^{yw} \text{d}^4k,\label{eq:second_chern}
\end{align}
where $ \epsilon_{\alpha \beta \gamma \delta }$ is the 4D Levi-Civita symbol and where we have written out the antisymmetric wedge product $\Omega \wedge \Omega$ in components for clarity. Now the integral is taken over the first (magnetic) BZ in 4D, which we denote by $\mathbb{T}^4$. As can be seen, the 2CN is a genuine four-dimensional invariant, which vanishes in lower-dimensional systems due to the 4D Levi-Civita symbol. Physically, the 2CN underlies the quantization of current in the 4D quantum Hall effect as will be introduced below.

We emphasise that in the above definitions we have explicitly restricted ourselves to a single isolated energy band. More generally, there may be an isolated set of bands amongst which there are degeneracies, such as for a spin-1/2 particle in the presence of time-reversal symmetry. Then the components of the Berry curvature $\Omega^{\mu \nu}$ are themselves matrices, with indices running over the set of bands. The definitions of the 1CN (\ref{eq:first_chern}) and the 2CN (\ref{eq:second_chern}) can then be generalised to the integral of a matrix-trace over $\Omega$ and $\Omega \wedge \Omega$ respectively. We note that while the 1CN so-defined always vanishes without time-reversal symmetry breaking, there can be non-zero 2CNs also in a time-reversal invariant system in the presence of an SU(2) gauge field; indeed, this was the context in which the 4D quantum Hall effect was originally proposed~\cite{IQHE4D} and studied~\cite{QiZhang,LiZhang2012,Li:2012, Li:2013, Edge:2012,Kraus:2013}. 

\subsection{Semiclassical Equations of Motion} \label{sec:semiclass}

Having introduced the geometrical and topological properties of the underlying energy bands, we now review the motion of a wave packet prepared in a given Bloch band $\mathcal{E}({\bs k})$ and subject to perturbing electromagnetic fields~\cite{chang1, chang2, XiaoPol, Gao, Gao2}.

\subsubsection{Introducing the perturbing fields}\label{sect:perturbing}

We consider particles of charge $-e$ moving in the presence of two families of fields: (1) the ``intrinsic" fields $\mathcal{B}_{\mu \nu}$  generating the band structure $\mathcal{E}({\bs k})$ under scrutiny, and (2) the perturbing ``extrinsic" fields in response to which transport is analyzed. While the intrinsic fields $ \mathcal{B}_{\mu \nu}$ do not need to be specified at this stage [their effects are entirely captured by the dispersion $\mathcal{E}({\bs k})$ and Berry curvature $\Omega^{\mu \nu} (\bs k)$ of the band], the perturbing fields will be taken in the form of a weak electric field $\bs E= E_{\mu} \bs{e}^{\mu}$ and a weak magnetic field strength $B_{\mu \nu}\!=\!\partial_{\mu}A_{\nu}\!-\!\partial_{\nu}A_{\mu}$, where $\bs A= A_{\mu} \bs{e}^{\mu}$ denotes the electromagnetic vector potential. We assume that these weak external electromagnetic fields are both time-independent and spatially-uniform.  In the following, we also set Planck's constant $\hbar\!=\!1$ and the elementary charge $e\!=\!1$, such that $h/e^2\!=\!2 \pi$. We note that the discussion that follows is general and also directly applies to neutral particles subject to synthetic gauge fields~\cite{Dalibard2011, Goldman:2014bv}. 

Finally, we point out that, while the field $B_{\mu \nu}$ should be weak enough for the following perturbative analysis to be valid, the partition of external magnetic fields into intrinsic [$ \mathcal{B}_{\mu \nu}$] and perturbing [$B_{\mu \nu}$] components is  somewhat arbitrary; this aspect will be illustrated in Sections~\ref{sec:2D}-\ref{sec:4D}. 

\subsubsection{Equations of Motion}

In a semiclassical approach, motion is assumed to be adiabatic with respect to a manifold of states, such that a wave packet can be constructed out of this manifold at all times. The wave packet is chosen to have a well-defined center of mass at position $\bs r_c\!=\! r_c^{\mu} \bs{e}_{\mu}$ and quasi-momentum $\bs k^c\!=\! k^c_{\mu} \bs{e}^{\mu}$. The appropriate manifold to use for the wave packet construction depends on the strength of the applied electromagnetic fields. To see this, the full quantum Hamiltonian, including the perturbing electromagnetic fields, can be expanded around ${\bs r}_c$ as~\cite{chang1, chang2, XiaoPol, Gao, Gao2, Xiao2010}
\begin{align}
\hat{H} \approx \hat{H}_c + \hat{H}' + \hat{H}''+... \,,
\end{align}
where $\hat{H}_c$ is the full Hamiltonian evaluated at the center-of-mass position, and $\hat{H}'$ ($\hat{H}''$) are first-(second-)order gradient corrections in the electromagnetic fields. 

When the external fields are weak, the wave packet moves adiabatically with respect to the eigenstates of an isolated energy band $\mathcal{E}({\bs k})$ of $\hat{H}_c$, corresponding to the original Bloch states $| {u}_{\bs k} \rangle $ introduced above, up to a phase~\cite{chang1, chang2}. The resulting semiclassical equations of motion are then consistent up to first order in the perturbing electromagnetic fields; this is sufficient to capture the physics of the 2D quantum Hall effect for a filled band. To extend this validity up to second order, so that we may also capture the (nonlinear) 4D quantum Hall response, the wave packet should be constructed out of the perturbed states $| \tilde{u}_{\bs k}  \rangle = | {u}_{\bs k}  \rangle + | \tilde{u}_{\bs k} ' \rangle $ where $ | \tilde{u}_{\bs k} ' \rangle$ are the first-order band-mixing corrections from $\hat{H}'$~\cite{XiaoPol, Gao, Gao2}. These perturbed states have geometrical properties encoded in modified Berry curvature components~\cite{Gao}
\begin{align}
\tilde{\Omega}^{\mu \nu} &= \Omega^{\mu \nu} +  \Omega_{1}^{\mu \nu} \,, \nonumber \\
 \Omega_1^{\mu \nu} &=  \partial_{k_\mu} \mathcal{A}'_{k_\nu }  -  \partial_{k_\nu} \mathcal{A}'_{k_\mu} \,, \label{eq:berry}
\end{align}
where $\mathcal{A}'_{k_\mu } = i \langle u |  \partial_{k_\mu } | u' \rangle + \mbox{c.c.}$ is the first-order correction to the Berry connection. As derived in Ref.~\cite{Gao}, the semiclassical equations valid up to second order in dimensions $d \ge 2$ are
\begin{align}
\dot{r}^\mu ({\bs k}) &= \frac{\partial \tilde{\mathcal{E}} (\bs k)}{\partial k_\mu } - \dot{k}_\nu \tilde{\Omega}^{\mu \nu}  (\bs k), \label{eq:semir} \\
\dot{k}_\mu &= - E_\mu - \dot{r}^\nu  B_{\mu \nu} \,, \label{eq:semik}
\end{align}
where we have dropped the subscript $c$ from the center-of-mass position and quasi-momentum. Here, the second-order wave packet energy $\tilde{\mathcal{E}}({\bs k})$ contains the unperturbed Bloch band $\mathcal{E}({\bs k})$ plus corrections from the perturbing terms $\hat{H}' + \hat{H}''$~\cite{Gao, Gao2}. These equations can be combined repeatedly to give the mean velocity of the wave packet~\cite{4Datoms:2015}
\begin{align}
\dot{r}^\mu 
&\approx \resizebox{.85\hsize}{!}{$\displaystyle{\frac{\partial \tilde{\mathcal{E}}}{\partial k_\mu }+  E_\nu  \tilde{\Omega}^{\mu \nu}+ \left( \frac{\partial\tilde{\mathcal{E}}}{\partial k_\gamma }  +  E_\delta  \tilde{\Omega}^{\gamma \delta} +   \frac{\partial \tilde{\mathcal{E}}}{\partial k_\alpha }   B_{ \delta \alpha}  \tilde{\Omega}^{\gamma \delta} 
\right)B_{ \nu \gamma}  \tilde{\Omega}^{\mu \nu}}$} , \label{eq:meanv}
\end{align}
where terms above second-order in the perturbing electromagnetic fields are neglected. As can be seen, there are many terms in this expression; these will lead both to Bloch oscillations and unquantized anomalous Hall effects for a wave packet~\cite{Gao, Gao2}. 

In order to reveal the topological quantum Hall effects, one has to compute the total current density $j^{\mu} $ associated with an entirely populated band. This can be obtained using the mean velocity in Eq.~\eqref{eq:meanv} and summing over all the states located in the band, 
\be
j^{\mu} =\frac{1}{L^d} \sum_{\bs k} \rho(\bs k) \, \dot{r}^\mu (\bs k),\label{totalcurrent}
\ee
where $L^d$ is the volume of the system of dimension $d$, and where $\rho(\bs k)$ is the distribution function of particles within the band. 
In general, converting the sum over momentum states in Eq.~\eqref{totalcurrent} into an integral over the BZ is subtle~\cite{Xiao2010}: in the semiclassical limit, this operation can be performed through the so-called modified density of states $ D({\bs r}, {\bs k})$,
\begin{align}
\frac{1}{L^d} \sum_{\bs k} \rho(\bs k) \rightarrow \int_{\mathbb{T}^d} \text{d}^d k D({\bs r}, {\bs k}) \rho(\bs k) \,,\label{eq:sum}
\end{align}
where the integration is performed over the first (magnetic) Brillouin zone ($\mathbb{T}^d$). Notably, the explicit expression for the modified density of states relies on an interesting interplay between the perturbing magnetic field $B_{\mu \nu}$ and the Berry curvature of the band $\Omega^{\mu \nu}$. 

\subsection{Modified Density of States} \label{sec:modified}

When all magnetic field effects are included intrinsically into the band structure, the phase-space density of states $D({\bs r}, {\bs k})$ is a simple constant factor of $1/h^d\!=\!1/ (2 \pi)^d$. This is also the case when all magnetic fields are treated extrinsically as perturbations acting on systems with zero Berry curvature.
The fact that the density of states remains constant is guaranteed classically by Liouville's theorem, which states that the phase-space volume element is conserved under time evolution~\cite{Xiao_ED}. However, Liouville's theorem holds for the volume element $\Delta V = \Delta {\bs R} \Delta {\bs K}$ associated with the {\it canonical} position ${\bs R}$ and momentum ${\bs K}$, while the semiclassical treatment above is for the {\it physical} position ${\bs r}$ and momentum ${\bs k}$~\cite{Xiao_ED, Bliokh, Duval}. 

To see how canonical and physical variables are related, we consider three cases in turn. Firstly, if all magnetic effects are treated extrinsically, i.e. are not incorporated into the band structure, and if the unperturbed band structure has a trivial geometry (vanishing Berry curvature), the particle moves in the presence of a magnetic vector potential but a trivial Berry connection. Then the physical momentum is modified by the minimal (or Peierls) substitution ${\bs k} = {\bs K} - {\bs A} ({\bs r})$, while the physical and canonical positions are equivalent~\cite{Landau}. 

Secondly, if all magnetic effects are treated intrinsically, i.e. are included in the band structure, the particle moves in the presence of a nontrivial Berry connection but no (additional) magnetic vector potential. In this case, the physical position is ${\bs r} = {\bs R} + { \mathcal{A}} ({\bs k})$, while the physical and canonical momenta are equivalent~\cite{Adams, Nagaosa, Xiao2010}. The symmetry between these substitutions can be understood as the Berry curvature acting like a magnetic field in momentum space~\cite{Nagaosa, Bliokh_mag, Price:2014}, i.e. the Berry connection $\mathcal{A}$ and magnetic vector potential ${\bs A}$ are dual. 

Thirdly, if, as we consider here, some magnetic fields are treated intrinsically and some extrinsically, the particle experiences both a Berry connection and a magnetic vector potential. Then neither the physical position nor momentum remain equal to their canonical counterparts, and generalised Peierls substitutions are required~\cite{Gao, Xiao2010, Bliokh, Gosselin}. The modified density of states $D({\bs r}, {\bs k})$ can then be understood as the usual phase-space density of states $1/ (2 \pi)^d$ multiplied by the Jacobian of the transformation from the canonical to physical variables~\cite{Duval, Bliokh, Gosselin, Gosselin2}. For dimensions up to $d=4$, the modified density of states, valid to second-order in the external fields, is given by~\cite{4Datoms:2015}
\begin{align}
D({\bs r}, {\bs k}) = \frac{1}{(2\pi)^d} & \left[  1 + \frac{1}{2}  B_{\mu \nu} \tilde{\Omega}^{\mu \nu}  \right. +\frac{1}{64} \left( \varepsilon^{\alpha \beta \gamma \delta } B_{\alpha \beta }B_{\gamma \delta }\right)  \nonumber \\ 
&  \left. \times \left( \varepsilon_{\mu \nu \lambda \rho} {\Omega}^{\mu \nu }\Omega^{\lambda \rho }\right) \right]  \,, \label{eq:dos}
\end{align}
where, thanks to the Levi-Civita symbol, the last term vanishes in fewer than four dimensions~\cite{Xiao_ED, Duval, Bliokh, Gosselin}. Since our analysis is restricted to second-order, the last term in Eq.~\eqref{eq:dos} only involves the zeroth-order components of the Berry curvature ${\Omega}^{\mu \nu }$. In contrast, the first correction to the density of states, which is linear in the perturbing magnetic field $B_{\mu \nu}$, involves the first-order corrections to the curvature through $\tilde{\Omega}^{\mu \nu}$. Importantly, the correction $\Omega_1^{\mu \nu}$, introduced in Eq.~\eqref{eq:berry}, necessarily vanishes upon integration over the BZ as it is the curl of $\mathcal{A}'_{k_\mu }$, which is gauge-invariant and periodic in the BZ~\cite{Gao}. Hence, one can safely substitute $\tilde{\Omega}^{\mu \nu} \rightarrow \Omega^{\mu \nu}$ in Eq.~\eqref{eq:dos} whenever one considers a uniformly filled band. 

\subsection{Semiclassical Current Density} \label{sec:current}

The modified density of states \eqref{eq:dos} can be combined with the mean velocity \eqref{eq:meanv} to evaluate the semiclassical current density of a band filled with a given distribution of particles $\rho(\bs k)$. In standard quantum Hall systems, one typically considers an isolated band that is completely filled with spinless fermions, in which case $\rho(\bs k)\!=\!1$. The more general uniformly-populated-band situation [i.e. $\rho(\bs k)\!=\! \rho$] is also particularly relevant to cold-atom and photonics experiments (see Sections \ref{sect:c.m._atom}-\ref{sect:c.m._photon}). However, as the current density $j^{\mu}$ for this configuration can be simply obtained from the completely-filled-band result through the substitution $j^\mu (\rho)\!\rightarrow\! \rho j^\mu (\rho\!=\!1)$ [see Eq.~\eqref{totalcurrent}], we focus without loss of generality on the completely filled-band case. 

Then the semiclassical current density is %
\begin{align}
j^\mu 
 \approx& \int_{\mathbb{T}^d} \frac{ \text{d}^d k }{(2\pi)^d} 
  \left[   E_\nu  \tilde{\Omega}^{\mu \nu} +
  E_\delta \Omega^{\gamma \delta}B_{ \nu \gamma} \Omega^{\mu \nu} +  \frac{1}{2}E_\nu  {\Omega}^{\mu \nu} B_{\delta \gamma}  {\Omega}^{\delta \gamma} \right. \nonumber \\
  &+ \frac{\partial \tilde{\mathcal{E}}}{\partial k_\mu }+
    \frac{\partial \tilde{\mathcal{E}}}{\partial k_\gamma }  B_{ \nu \gamma} \tilde{\Omega}^{\mu \nu} + \frac{1}{2} \frac{\partial \tilde{\mathcal{E}}}{\partial k_\mu }B_{\gamma \nu}  \tilde{\Omega}^{\gamma \nu}     
  \nonumber \\ 
 &+  \left( \frac{\partial \mathcal{E}}{\partial k_\alpha } B_{ \delta \alpha} \Omega^{\gamma \delta} 
+ \frac{1}{2}\frac{\partial \mathcal{E}}{\partial k_\gamma }  B_{\delta \alpha}  {\Omega}^{\delta \alpha}  \right) B_{ \nu \gamma} \Omega^{\mu \nu}
  \nonumber \\ &+ \left.
  \frac{1}{64}  \frac{\partial \mathcal{E}}{\partial k_\mu }( \varepsilon^{\alpha \beta \gamma \delta } B_{\alpha \beta }B_{\gamma \delta} )( \varepsilon_{\xi \nu \lambda \rho} {{\Omega}}^{\xi \nu }{\Omega}^{ \lambda \rho} )  \right] \,,\label{eq:current}
 \end{align}
where whenever needed $\Omega$ replaces $ \tilde{\Omega}$ and $\mathcal{E}$ replaces $ \tilde{\mathcal{E}}$ to keep only terms up to second-order in the external fields. Expression (\ref{eq:current}) can be substantially simplified; firstly, terms on the third and fourth lines sum to zero due to the antisymmetry of the magnetic field strength and Berry curvature~\cite{4Datoms:2015}. As this cancellation is by symmetry, it holds also for a band with arbitrary filling. Secondly, it can be shown that terms on the second line vanish upon integration over the BZ, using, as needed, the periodicity of the corrected energy dispersion $\tilde{\mathcal{E}}$ and the Bianchi identity for the antisymmetric Berry curvature~\cite{XiaoPol}. This leaves only the first line which we rewrite as
\begin{align}
j^\mu
   &= E_\nu  \int_{\mathbb{T}^d}
 {\Omega}^{\mu \nu} \frac{\text{d}^d k }{(2 \pi)^d}  \nonumber \\
 &+   \varepsilon^{\mu \alpha \beta  \nu}  \frac{1}{8} E_\nu B_{ \alpha \beta} 
 \int_{\mathbb{T}^d}  \epsilon_{\gamma \delta \lambda \rho } \Omega^{\gamma \delta}  \Omega^{\lambda \rho}   \frac{\text{d}^d k }{(2 \pi)^d}, \label{eq:main}
 \end{align}
where we have used that the Berry curvature correction $\Omega_1 ({\bs k})$ vanishes upon integration over the BZ as commented above~\cite{Gao}. We note that this expression does not depend on the corrections to the energy and Berry curvature appearing in the full second-order semiclassical equations (\ref{eq:semir}) and (\ref{eq:semik}); these would however play a role in the dynamics of bands with non-uniform partial fillings, which we do not discuss further here.

It is important to note, additionally, that the inclusion of the modified density of states in calculating the current density~\eqref{eq:current} implies that the band is filled in the presence of the perturbing magnetic field  $B_{\mu \nu}$. The experimental implications of this will be discussed further in Section~\ref{sec:experimental}. We also point out that the perturbing magnetic field can potentially split the unperturbed Bloch band under scrutiny $\mathcal{E}({\bs k})$ into a set of subbands, in which case the filling condition discussed above should apply to the set of subbands.

\subsubsection{Linear vs nonlinear responses}

The first term in Eq.~\eqref{eq:main} can be nonzero for a system with two or more dimensions; this is the famous quantum Hall current response which is linear in the applied electric field. This is independent of any magnetic perturbations and, in 2D, is directly proportional to the 1CN in Eq.~\eqref{eq:first_chern}. The second term, conversely, may be nonzero only for systems with four or more dimensions due to the Levi-Civita symbols; it corresponds to a nonlinear current response  as it depends on both the applied electric field and the magnetic perturbing field. In 4D, the nonlinear term is directly proportional to the 2CN in Eq.~\eqref{eq:second_chern}. While we have  truncated the perturbative treatment at second-order in the applied fields, at each higher order there can be an additional quantum Hall response for systems with increasing even dimensions (potentially involving higher-dimensional topological invariants).  

\subsubsection{External field partitionment and the current response} \label{sec:external}

We observe that the current response $j^\mu$ in Eq.~\eqref{eq:main} should not depend on the arbitrary partition of external magnetic fields into intrinsic [$ \mathcal{B}_{\mu \nu}$] and perturbing [$B_{\mu \nu}$] components, see discussion in Section \ref{sect:perturbing}. In particular, one is formally allowed to include all external magnetic fields into the intrinsic component, which  in our framework is directly realized through the substitutions 
\be
\mathcal{B}_{\mu \nu}\!\rightarrow \mathcal{B}_{\mu \nu}+B_{\mu \nu}, \quad B_{\mu \nu}\!\rightarrow 0. 
\ee 
In this singular picture, the intrinsic Berry curvature  now depends on the included weak field components, $\Omega\!\rightarrow\!\Omega (B_{ \alpha \beta})$,  and the transport equation in Eq.~\eqref{eq:main} becomes
\begin{align}
j^\mu
   &= E_\nu  \int_{\mathbb{T}^d}
 {\Omega}^{\mu \nu} (\bs k; B_{ \alpha \beta}) \frac{\text{d}^d k }{(2 \pi)^d} ,\label{eq:mainbis}
 \end{align}
where the linear and non-linear responses are now mingled. We note that the area of the magnetic Brillouin zone $\mathbb{T}^d$ over which the integration is performed now also depends on the included magnetic perturbation, $A_{\text{MBZ}}\!=\!A_{\text{MBZ}}(B_{ \alpha \beta})$. Moreover, including the perturbation $B_{ \alpha \beta}$ within the band structure potentially leads to a band splitting [see e.g.~Fig.~\ref{Fig_1}], in which case a trace should be performed over the matrix-valued Berry curvature in Eq.~\eqref{eq:mainbis}.

In this picture, the relation between the current responses and the Chern numbers of the underlying band structure is obscured, and so is the quantization of (Hall) transport coefficients. This drawback is particularly well illustrated in the case of time-reversal invariant 4D systems subject to time-reversal-breaking perturbations \cite{IQHE4D,QiZhang,LiZhang2012,Li:2012, Li:2013, Edge:2012,Kraus:2013}: the connection between the 4D-QH current response and the 2CN of the bands is only made clear when treating all external U$(1)$ magnetic fields as perturbing components (i.e.~by working in a picture where the unperturbed system is time-reversal invariant). 

These observations highlight the fact that, when all magnetic fields are present at all times, the external field partition, although arbitrary and formal, could be chosen based on theoretical convenience (see Sections~\ref{sec:2D}-\ref{sec:4D} for illustrations). However, in some experimental systems, as we discuss in Section~\ref{sec:experimental}, it is possible to ramp up magnetic perturbations after the initial preparation of the energy bands, in which case the partionment of the field into intrinsic and extrinsic components follows naturally.

\subsection{Center-of-Mass Observables} \label{sec:c.m.}

While current or voltage measurements have long been used to study the quantum Hall effect in solid-state systems, other measurements are easier to make in the analogue quantum Hall systems which are currently of great experimental interest~\cite{aidelsburger2013,miyake2013,jotzu2014,Aidelsburger:2015, Mancini:2015,Stuhl:2015, Hafezi,Hafezi:2014}. In particular, key experimental observables in ultracold atomic gases and photonics can be related to center-of-mass (c.m.) behaviour, e.g. the center-of-mass motion of an atomic cloud~\eqref{eq:atomcom}. Such c.m. observables depend generally not just on the quantum Hall current (\ref{eq:main}), but also on the particle density $n$, which is calculated semiclassically as: 
\begin{eqnarray}
n&=&   \int_{\mathbb{T}^d}  \text{d}^dk \,  D({\bs r}, {\bs k}) \rho(\bs k) .\label{eq:densityrealistic}
\end{eqnarray}
As the modified density of states $ D({\bs r}, {\bs k})$ directly depends on the Berry curvature $\Omega$, see Eq.~\eqref{eq:dos}, the particle density of a filled band potentially contains information about the topology of the filled band [see Sections \ref{sect:diophantine} and \ref{section:case3} below]. This has important implications for experiments that extract Chern numbers from the measurements of c.m. observables.   

\subsubsection{Center-of-mass drift in cold atoms}  \label{sect:c.m._atom}

In ultracold atoms, as introduced above, a natural observable is the center-of-mass motion of a cloud, which can be extracted from in-situ density measurements as in the experiment of Ref.~\cite{Aidelsburger:2015}. As shown in Eq.~\eqref{eq:atomcom}, the c.m. velocity for a cloud of atoms is given by ${\bs v}_\text{c.m.} \!=\!  {{\bs j}}/{n}$, and so is influenced by both the (quantum Hall) current response (\ref{eq:main}) and the particle density $n$~\eqref{eq:densityrealistic}, with important consequences discussed below.

We also point out that the above semiclassical analysis equally applies to systems of non-interacting fermions and bosons, as it only relies on the distribution function of particles within the band $\rho(\bs k)$. Importantly, when a Bloch band is uniformly populated, $\rho(\bs k)\!=\! \rho$, the c.m. observables such as in  Eq.~\eqref{eq:atomcom} become independent of the band filling factor $\rho$. This is because both the current density $j^{\mu}$ from Eq.~\eqref{totalcurrent} and the particle density $n$ in Eq.~\eqref{eq:densityrealistic} are directly proportional to the band filling factor $\rho$. 

The uniformly-populated-band situation [$\rho(\bs k)\!=\! \rho$] is particularly relevant to cold-atom experiments, whenever the temperature $T$ is large compared to the bandwidth $W$ of the lowest-energy band, but small (or of the order) of the band gap, $W\! \ll \! k_{\text{B}}T \! \ll \! \Delta$. This typically occurs when the lowest Bloch band displays a large flatness ratio $\Delta/W\!\gg\! 1$, as was recently demonstrated in the Munich experiment~\cite{Aidelsburger:2015} through band-mapping. In particular, this indicates that atomic transport experiments based on center-of-mass responses could be equally performed using (thermal) Bose or Fermi gases; this is in sharp contrast to measurements based on current densities, where the filling factor $\rho$ should be independently measured. 

\subsubsection{Center-of-mass displacement in photonics} \label{sect:c.m._photon}

In photonics, an optical analogue of the quantum Hall effect could be measured in the displacement of the center-of-mass of the photon steady-state in a driven-dissipative system of coupled photonic cavities~\cite{Ozawa:2014}. In such a system, a continuous-wave laser can pump light resonantly with a given isolated energy band, while the photon loss rate $\gamma$ in the lattice is chosen such that $W \!\ll\! \gamma \!\ll\! \Delta$, where $W $ is the band-width of the chosen energy band, and $\Delta $ is the band-gap to the nearest energy band. In this regime, for sufficiently long times, the light reaches a non-equilibrium steady-state in which the losses lead to an approximately uniform population of the single energy band. Hence the center-of-mass displacement of the photon steady state for a square (hypercubic) lattice can be expressed as~\cite{Ozawa:2014, 4Dphotons:2015}
\begin{eqnarray}
\langle {\bs r}_\text{photon} \rangle  = \frac{ \sum_{\bs r} {\bs r} |a_{\bs r}|^2}{ \sum_{\bs r} |a_{\bs r}|^2} \approx \frac{{\bs j}}{ \gamma  n  } \label{eq:photon}
\end{eqnarray}
where $a_{\bs r}$ is the expectation value of the photon field in the cavity labelled by ${\bs r}$, the d-dimensional position index, and where $n$ is again the particle density of the filled band~\eqref{eq:densityrealistic}.  This idea is not restricted to optics but an analogous effect could also be observed in any driven-dissipative system of coupled-classical harmonic oscillators simulating quantum Hall physics~\cite{Salerno:2015}. 

Due to the similarities between Eqs.~(\ref{eq:atomcom}) and (\ref{eq:photon}), our discussion in the following sections focuses on the center-of-mass velocity for ultracold atoms but has important experimental implications also in photonics or even in classical mechanical systems. However, we note that the displacement of the steady state in Eq.~(\ref{eq:photon}) is the leading-order term to which there are corrections that, for example, do not depend on the loss-rate, and which should be included when modeling realistic experiments~\cite{Ozawa:2014, 4Dphotons:2015}. 

\section{Magnetic Perturbations and the 2D Quantum Hall effect}  \label{sec:2D}

To emphasise the differences between the current density and center-of-mass response, we first discuss the important case of a two-dimensional quantum Hall system in the $x-y$ plane. As defined in Eq.~\eqref{eq:first_chern}, the integral of the Berry curvature over the whole 2D BZ gives the topological first Chern number $\nu_1^{xy}$ of the band. In the following, we focus on the 2D Harper-Hofstadter model, introduced below, as a concrete example of a system with energy bands that have nontrivial 1CNs. However, we note that it is straightforward to extend our discussion to other topologically nontrivial 2D models as required. 

\subsection{2D Harper-Hofstadter Model}\label{section:harper}

The 2D Harper-Hofstadter (HH) model is a seminal lattice model for studying the quantum Hall effect that was originally developed to describe a charged particle hopping on a 2D tight-binding square lattice in the presence of a uniform perpendicular magnetic field $\bs{B}\!=\!B \bs{e}_z$~\cite{Hofstadter}. The Hamiltonian is given by:
\begin{eqnarray}
\hat H=-J &  \sum_{\bs r} \left(c^{\dagger}_{\bs r + a\bs e_x}c_{\bs r}+ e^{i 2 \pi \Phi x/a}  c^{\dagger}_{\bs r + a\bs e_y}c_{\bs r} + \text{h.c.}\right), \qquad
\end{eqnarray}
where $c^{\dagger}_{\bs r}$ creates a fermion at lattice site $\bs r\!=\!(x,y)$, $a$ is the lattice spacing, $J$ is the hopping amplitude, and $\Phi\!=\! - a^2 B / {2\pi}$ is the magnetic flux per plaquette in units of the flux quantum. Here we have chosen the magnetic vector potential in the Landau gauge such that the hopping along $\hat{y}$ is modified by complex spatially-dependent Peierls phase factors, while the hopping along $\hat{x}$ is unaffected by them. 

A rational magnetic flux per plaquette $\Phi = p/q$, where $p$ and $q$ are coprime integers, can be directly incorporated into magnetic Bloch states~\cite{Avron1988}. These are arranged into $q$ energy bands described by a bandstructure $\mathcal{E}_{n}(k_x, k_y)$ in the so-called magnetic Brillouin zone, see Fig.~\ref{Fig_1} (a). The magnetic Brillouin zone is defined by the magnetic translational symmetry of the HH model, and is a factor of $q$ smaller than the original BZ, having an area $A_{\text{MBZ}} = (2 \pi)^2 / q a^2$. Due to the incorporated magnetic flux, the eigenstates in the bands have nontrivial Berry curvatures and nonzero 1CNs. For suitable values of the flux (e.g. $\Phi=1/q$),  the energy spectrum has a non-degenerate lowest band which is well-separated from other bands, to which the above semiclassical approach can be directly applied. 

\subsubsection{Physical realizations of the HH model}

The HH model is a directly experimentally-relevant model in materials, where it has been realised for electronic transport in graphene placed on boron nitride substrates~\cite{Dean, Ponomarenko}. Furthermore, the HH model has recently been generated in a wide-variety of analogue systems with neutral particles, where the spatially-dependent complex (Peierls) phase-factors in the tunneling matrix elements are carefully engineered. In ultracold gases, for example, the HH model has been realised by trapping atoms in a 2D optical lattice, and then combining a superlattice (or a Wannier-Stark ladder) along one direction with a resonant time-modulation of the optical-lattice potential~\cite{aidelsburger2013,miyake2013,Aidelsburger:2015}. In an alternative approach, atoms were trapped in a one-dimensional optical lattice while two-photon Raman couplings induced transitions between different internal atomic states~\cite{Mancini:2015,Stuhl:2015}. In this set-up, the internal states could then be viewed as sites along an extra ``synthetic" dimension~\cite{Boada2012,Celi:2014}, meaning that the atoms moved in an effective 2D lattice. By controlling the spatial dependence of the Raman wave-vector, the experiments were able to implement complex hopping phase-factors along this synthetic dimension, and hence realise HH physics. 

For photons, the HH model has been experimentally simulated in an array of silicon ring resonators~\cite{Hafezi}, where link resonators were used to introduce artificial Peierls phase-factors. The concept of synthetic dimensions can also be extended to photonics~\cite{peano:2015}; it has been proposed to realise a 2D HH model in either a one-dimensional array of optical cavities where different angular momentum modes are coupled by spatial light modulators~\cite{Luo:2015} or in a one-dimensional array of ring resonators, where the modes are coupled via an external time-dependent modulation~\cite{4Dphotons:2015}. The HH model has also been implemented by controlling inter-site couplings in 2D arrays of circuit elements~\cite{Simon, Albert} and classical pendula~\cite{Huber}. Additionally, proposals exist for systems of periodically-modulated classical coupled harmonic oscillators~\cite{Salerno:2015}. Hence, a full understanding of center-of-mass observables in this model can have important and direct applications in many current experiments. 

\subsection{2D Quantum Hall Response} 

To study the quantum Hall response in two dimensions, we consider an electric field applied along the $y$ direction $\bs E= E_{y} \bs{e}^{y}$ and a perturbing magnetic field $B_{x y} \ll B$. 
In experiments with neutral particles, such as those introduced above, these perturbing fields can also be imposed artificially. An electric field could correspond, for example, in ultracold gases to a linear gradient created either magnetically \cite{aidelsburger2013,miyake2013} or optically \cite{Aidelsburger:2015}. In a 2D array of coupled photonic cavities, it can be generated by a spatial gradient in the cavity size or temperature; see also Ref.~\cite{4Dphotons:2015} for synthetic electric fields acting along synthetic dimensions. 

The weak magnetic field $B_{x y}$ may also arise naturally in experiments whenever the realized flux $\Phi_{\text{expt}}\!=\! \Phi \!+\! \tilde{\Phi}$ slightly deviates from the desired (rational) value $\Phi\!=\!p/q$. In the following, we write the perturbing flux as
\be
\tilde \Phi\!=\! - a^2 B_{xy} / {2\pi}.\label{def:pertflux}
\ee
 In recent cold-atom experiments, for example, the flux was estimated to be ${\Phi}_{\text{expt}}\!\approx\! (1/4)\!\times\! 0.73(5)$ in \cite{aidelsburger2013}, ${\Phi}_{\text{expt}}\!\approx\! 0.185$ in \cite{Mancini:2015}, ${\Phi}_{\text{expt}}\!\approx\! 4/3$ in \cite{Stuhl:2015} and ${\Phi}_{\text{expt}}\!\approx\! 1/4$ in \cite{Aidelsburger:2015}. In photonic lattices, the flux realised in Ref.~\cite{Hafezi} was $\Phi_{\text{expt}} \approx 0.15$, although there was also an additional random variation in the flux over the lattice.

\subsubsection{Current Density}

Under the perturbing fields $\bs E$ and $B_{x y}$, the current density of the filled lowest band from Eq.~\eqref{eq:main} simply leads to Eq.~\eqref{eq:jx2dintro}, written out again here, now with our choice of units $e\!=\!\hbar\!=\!1$:
\begin{eqnarray}
j^{x} = \frac{E_y}{2 \pi} \nu_1^{xy} .\label{eq:jx2d}
\end{eqnarray} 
This is the 2D quantum Hall current response \cite{TKNN}, for which a weak perturbing magnetic field has no direct effect.  

\begin{figure*}[t!]
\includegraphics[width=18cm]{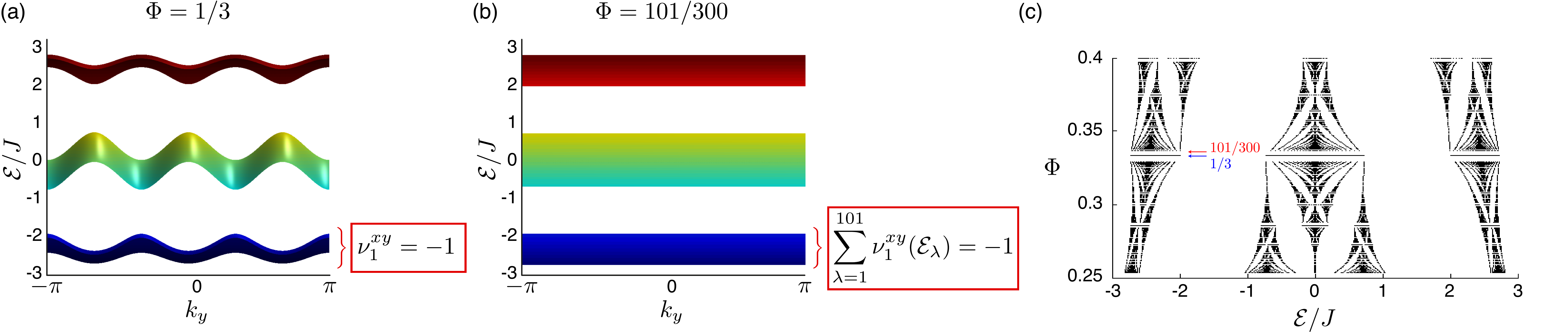}
\vspace{-0.cm} \caption{Energy spectrum $\mathcal{E}(k_x,k_y)$ of the 2D Harper-Hofstadter model~\cite{Hofstadter} for (a) $\Phi\!=\!1/3$ and (b) $\Phi\!=\!101/300$. 
For $\Phi\!=\!1/3$, the spectrum contains three well-separated bands, where the 1CN of the lowest band is indicated. Increasing the flux to $\Phi\!=\!101/300$ splits the lowest band into 101 subbands, while preserving the sum of 1CNs as shown. Due to the large number of bands in (b) compared to (a) it can be useful to instead view a system with total flux $\Phi\!=\!101/300$ as a system with a strong flux $\Phi\!=\!1/3$ and a weak perturbing flux $\tilde{\Phi}\!=\!1/300$. (c) Hofstadter butterfly (energy as a function of the magnetic flux $\Phi$) in the vicinity of the value $\Phi\!=\!1/3$. The lowest band at exactly $\Phi\!=\!1/3$ is shown to split into many subbands as the flux deviates from this ideal value, leading to complex (fractal) patterns in the $\mathcal{E}-\Phi$ plane.
}\label{Fig_1}
\end{figure*}

Let us comment further on why it may be convenient to separate the flux penetrating a 2D lattice into strong and weak components, i.e. $\Phi_{\text{tot}}\!=\!\Phi \!+\! \tilde{\Phi}$. Indeed,  imagine that a synthetic flux $\Phi_{\text{expt}}\!=\!101/300\!\approx\!0.34$ has been realized; the corresponding energy spectrum splits into three isolated sets of bands, each set being  associated with numerous extremely flat subbands as shown in Fig.~\ref{Fig_1}(b)-(c) (e.g. the lowest set is constituted of 101 subbands). 
The same system can be seen as a lattice pierced by a main flux $\Phi\!=\!1/3$, leading to three non-degenerate isolated bands [Fig.~\ref{Fig_1}(a)], which is then slightly perturbed by a very weak flux $\tilde{\Phi}\!=\!1/300$. This latter picture, which involves a small set of non-degenerate bands, significantly simplifies the analysis of the Hall current, which remains immune to weak perturbing fields, see Eq.~\eqref{eq:jx2d}. 

\subsubsection{Semiclassical particle density and the Diophantine equation}\label{sect:diophantine}

Unlike the current response, the density of particles in a filled band is highly sensitive to perturbing magnetic fields. Semiclassically this can be seen from Eq.~\eqref{eq:densityrealistic} calculated for a 2D system
\begin{eqnarray}
n &=& \int_{\mathbb{T}^2} \frac{\text{d}^2k}{(2\pi)^2}  \left( 1 +  B_{xy}{\Omega}^{xy} \right)  \nonumber \\
 &=& \frac{ A_{\text{MBZ}}}{(2 \pi)^2} +  \frac{ B_{xy} }{2\pi}   \nu_1^{xy}  , \label{eq:n1}
\end{eqnarray}
where we considered a filled lowest band, and where we used that first-order corrections to the Berry curvature, $\Omega_1$, vanish upon integration [see Eq.~\eqref{eq:dos} and discussion below]. The particle density therefore varies smoothly with a weak magnetic perturbing field in proportion to the 1CN of the filled band. This is as expected from the Streda-Widom formula~\cite{Widom, Streda, Smrcka}, which relates the Hall conductance to the variation of the particle density with respect to the magnetic field at fixed temperature and chemical potential. 

\begin{figure}
\includegraphics[width=9cm]{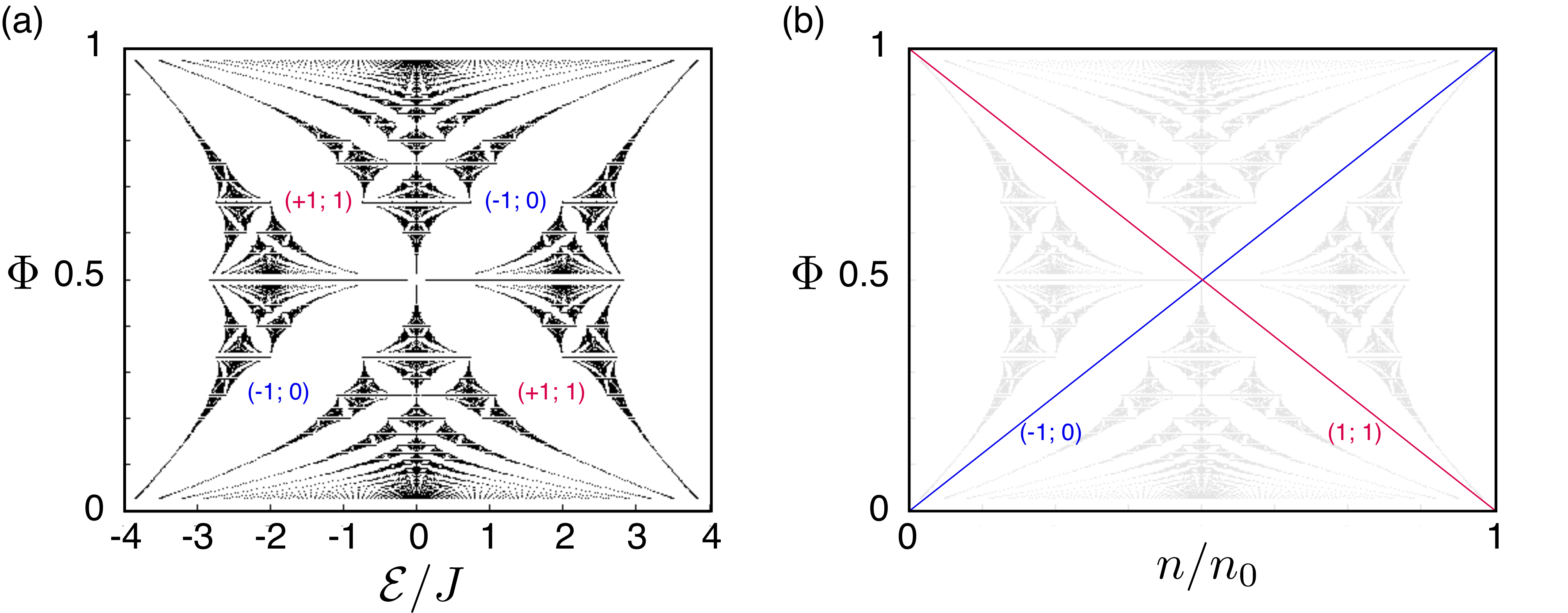}
\vspace{-0.cm} \caption{ (a) The Hofstadter butterfly: energy spectrum as a function of the magnetic flux $\Phi$. The main gaps are labelled by two integer $(t_r,s_r)$, which satisfy the Diophantine equation: $r\!=\!q s_r - p t_r$, where the integer $r$ denotes the $r$th gap and where $\Phi\!=\!p/q$. The integer $t_r$ is given by the sum of Chern numbers associated with the bands below the $r$th gap, and for the HH model was shown to satisfy $-q/2 < t_r < q/2$~\cite{Avron1988}.  Focusing on the first gap ($r\!=\!1$), this integer is simply given by the Chern number of the lowest band $t_1\!=\!\nu_1^{xy}$, see Eq.~\eqref{eq:diophantine}. (b) The Wannier diagram associated with the main gaps of the butterfly~\cite{Wannier}. In the $r$th gap, the reduced particle density satisfies the equation $n/n_0\!=\!s_r - \Phi t_r$, where the integers $(t_r,s_r)$ satisfy the aforementioned Diophantine equation.}
\label{Fig_wannier}
\end{figure}

As a concrete example of this physics, we consider the HH model introduced above, for which $A_{\text{MBZ}} = (2 \pi)^2 / q a^2$. Introducing the normalised particle density, we find [Eq.~\eqref{eq:n1}]
\begin{eqnarray}
\frac{n}{n_0}
 &=& \frac{1}{q }  - \tilde{\Phi}  \nu_1^{xy}, \label{eq:semiwannier}
\end{eqnarray} 
where $n_0 = 1 /a^2$ is the inverse of the unit cell area and $\tilde{\Phi}$ is the perturbing magnetic flux defined in Eq.~\eqref{def:pertflux}. This can be understood as the semiclassical derivation of the Wannier diagram~\cite{Wannier}: as illustrated in Fig.~\ref{Fig_wannier}, each gap in the energy spectrum of the HH model (the so-called Hofstadter butterfly \cite{Hofstadter}) can be described by a straight line on a plot of density versus applied magnetic flux~\cite{Wannier}. The slope of each line is given by minus the sum of first Chern numbers of the bands lying below the gap~\cite{dana:1985,Kohmoto:1989}. For the lowest gap, i.e. considering the lowest band, this line is simply given by~\cite{Kohmoto:1989,Dean}
\begin{eqnarray}
\frac{n}{n_0}
 &=& s - \Phi_{\text{tot}}  \nu_1^{xy}, \label{eq:wannier}
\end{eqnarray}
where $\Phi_{\text{tot}} = \Phi+ \tilde{\Phi}$ is the total magnetic flux per plaquette, and $s$ is an integer. For rational flux $\Phi_{\text{tot}} = \Phi = p/q$, the normalised particle density is simply ${n}/{n_0} = 1/ q$ as only one of $q$ bands is filled; this leads to the well-known Diophantine equation for the HH model~\cite{Wannier, TKNN}
\begin{eqnarray}
1 &=& q s - p \nu_1^{xy} .  \label{eq:diophantine}
\end{eqnarray}
Rearranging this as an equation for the integer $s$ and substituting this back into Eq.~\eqref{eq:wannier} then recovers the semiclassical result of Eq.~\eqref{eq:semiwannier}. 
While we have focused here on the HH model, we note that results such as Eqs.~\eqref{eq:wannier} \& \eqref{eq:diophantine} can be derived relying only on magnetic translational symmetry in a two-dimensional periodic potential~\cite{dana:1985}, while the semiclassical result \eqref{eq:n1} was derived for any system with a non-degenerate isolated energy band.

\subsubsection{Center-of-mass observables}\label{sect:2D_c.m.}

\begin{figure}[t!]
\includegraphics[width=7.5cm]{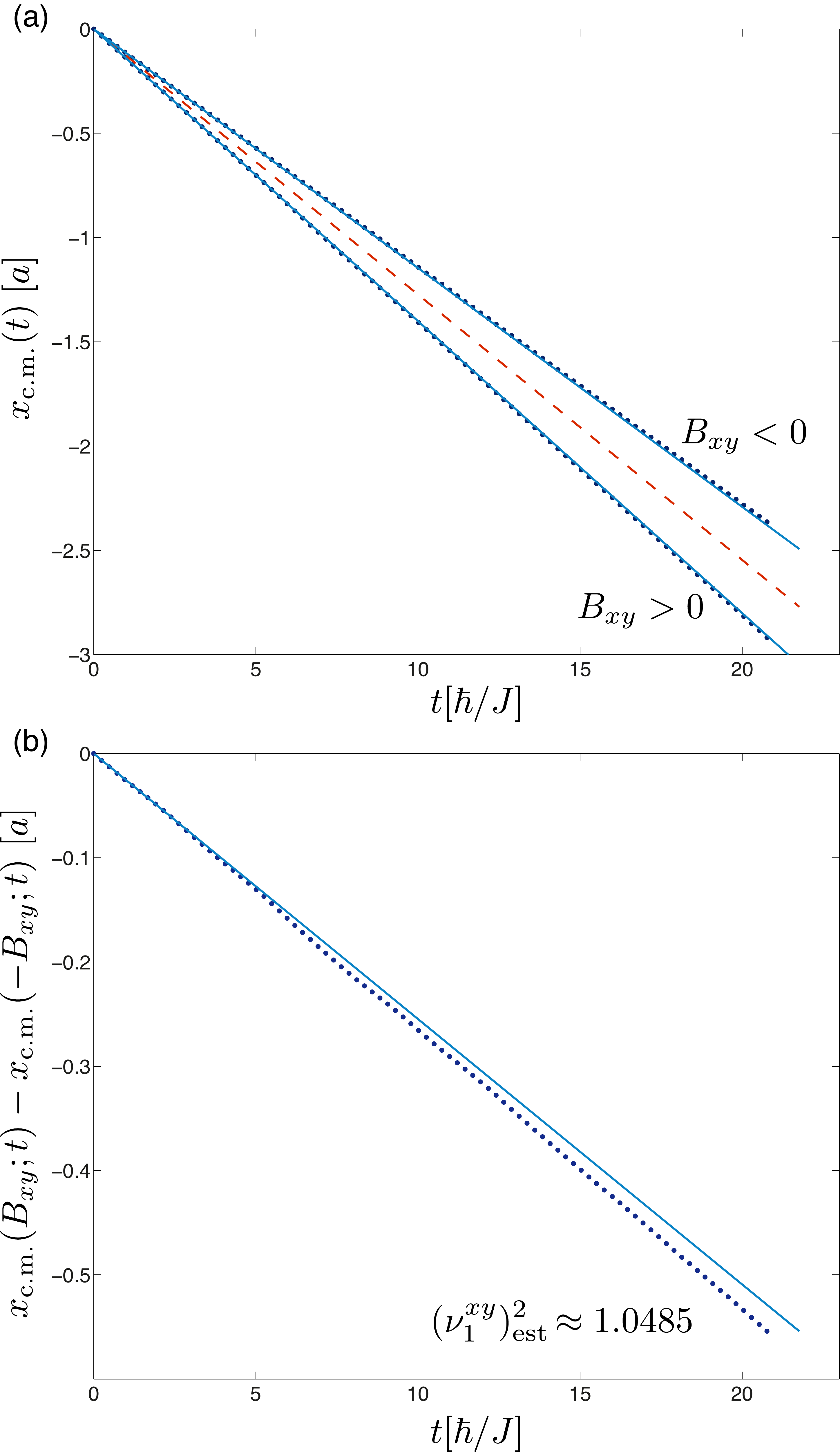}
\vspace{-0.cm} \caption{(a) Center-of-mass trajectories $x_{\text{c.m.}} (t)$ after ramping up the electric field to $E_y\!=\! 0.2 J/a$, in the presence of a strong flux $\Phi\!=\!1/4$ and a perturbing flux $\tilde{\Phi}\!=\!\pm 10\% \times \Phi$. The dark-blue dots are numerical simulations performed for a $200 \times 200$ HH lattice. The red dashed curve is the semiclassical analytical result in Eq.~\eqref{eq:taylor_result_2D} with $\nu^{xy}_1=-1$ when the nonlinear response due to the perturbing magnetic flux is neglected (i.e.~when disregarding the effects of the modified density of states). The light-blue solid curves are the corresponding analytical results when this response is also included. (b) The difference between the center-of-mass trajectories for positive and negative perturbing magnetic fluxes shown above in (a). The light-blue solid line is the analytical nonlinear response calculated from (\ref{eq:taylor_result_2D}). The dark-blue dots are numerical simulations, from which we extract an approximate $(\nu^{xy}_1)_{\rm{est}}^2$ as indicated on the plot.
The small deviation between the analytical and numerical results is due to approximation of Eq.~\eqref{eq:result_2D}
by  the Taylor series expansion in Eq.~\eqref{eq:taylor_result_2D}. Corrections to this approximation scale with the dimensionless parameter $\delta \!=\!( 2  \pi |B_{xy} \nu_1^{xy} |)/A_{\text{MBZ}}$; for the parameters shown here $\delta\!=\!0.10$. Residual deviations in the numerical results can be attributed to weak inter-band transitions and small inhomogeneities in the wave packet.}\label{Fig_2}
\end{figure}

As introduced above, center-of-mass observables can be important for experiments in analogue quantum Hall systems. Inputting the 2D current density (\ref{eq:jx2d}) and particle density (\ref{eq:n1}) into the c.m. velocity for an atomic cloud (\ref{eq:atomcom}), we find
\begin{eqnarray}
v_\text{c.m.}^x &=& \frac{j^x}{n} = 
\frac{ E_{y}  }{ \frac{A_{\text{MBZ}}}{2 \pi}+ B_{xy}  \nu_1^{x y }} \nu_1^{x y} . \label{eq:result_2D}
\end{eqnarray}
Neglecting any perturbing magnetic field $B_{xy}$, the c.m. velocity is directly proportional to the quantum Hall current response up to a simple multiplicative factor \cite{Dauphin:2013}. In such a configuration, the first Chern number has recently been experimentally extracted from a measurement of the center-of-mass drift of an ultracold cloud of atoms~\cite{Aidelsburger:2015}. However, when a perturbing magnetic field is present, the dependence on the first Chern number in Eq.~\eqref{eq:result_2D} is more involved. For a sufficiently weak additional magnetic field (i.e. $\vert 2  \pi B_{xy} \nu_1^{xy} \vert/A_{\text{MBZ}}\!\ll\! 1$), we can perform a Taylor expansion to write:
\begin{eqnarray}
v_\text{c.m.}^x &\approx \frac{  2\pi  }{ A_{\text{MBZ}}}  E_{y}\nu_1^{x y}  - \left (\frac{  2\pi   }{ A_{\text{MBZ}}} \right )^2   E_{y}B_{xy} \left(  \nu_1^{x y } \right)^2. \label{eq:taylor_result_2D}
\end{eqnarray}
As we see, even in 2D, an experiment could therefore measure a nonlinear response term in the center-of-mass drift, quantised in units of $\left(  \nu_1^{x y } \right)^2$. This second term can be isolated by performing differential measurements, reversing the sign of the perturbing magnetic field,  as discussed in Example II below and shown in Fig.~\ref{Fig_2}. We shall now detail two examples where this sensitivity to perturbing magnetic fields is distinctly manifested: 

\paragraph*{Example I:} Let us first illustrate the result in Eq.~\eqref{eq:result_2D} on a simple example. Consider a 2D lattice that is exactly pierced by a uniform flux $\Phi\!=\!1/5$. In this case, the lowest band of the spectrum is associated with a Chern number $\nu_1^{x y }\!=\!-1$, the area of the magnetic Brillouin zone is $A_{\text{MBZ}}\!=\!(2\pi/a)^2/5$, and the perturbing magnetic field is $B_{xy}\!=\!0$. According to Eq.~\eqref{eq:result_2D}, the c.m. velocity is thus given by $v_\text{c.m.}^x\!=\! (-5 a^2/2 \pi) E_y$.

Now, one can reinterpret this system as being a 2D lattice pierced by an intrinsic uniform flux $\Phi\!=\!1/4$,  which is perturbed by an extrinsic flux $\tilde{\Phi}\!=\!-1/20$. The total flux is then $\Phi_{\text{tot}}\!=\!\Phi\!+\tilde{\Phi}\!=\!1/5$, as defined above. The unperturbed spectrum, with flux $\Phi\!=\!1/4$, is still characterized by a lowest band with Chern number $\nu_1^{x y }\!=\!-1$, but now the area of the magnetic Brillouin zone is $A_{\text{MBZ}}\!=\!(2\pi/a)^2/4$ and the perturbing magnetic field is $B_{xy}\!=\!-2 \pi \tilde{\Phi}/a^2\!=\!\pi/10 a^2$. One readily verifies that Eq.~\eqref{eq:result_2D} yields $v_\text{c.m.}^x\!=\! (-5 a^2/2 \pi) E_y$, in agreement with the complementary picture above. Interestingly, this simple equivalence could not have been demonstrated without invoking the modified density of states in Eq.~\eqref{eq:n1}, and thus, its impact on c.m. observables. 

 \paragraph*{Example II:} Motivated by recent experiments in ultracold gases~\cite{aidelsburger2013, Aidelsburger:2015}, we consider a 2D lattice pierced by a uniform flux $\Phi\!=\!1/4$ with an additional uniform uncertainty in the flux of $\tilde{\Phi}=\pm 10\% \times \Phi$. To validate the above semiclassical results, we have numerically simulated the c.m. displacement $x_{\text{c.m.}}(\pm B_{xy}; t)$ of an ultracold cloud as shown in Fig.~\ref{Fig_2}. Numerically, the cloud is initially confined in the presence of all fluxes $\Phi_{\text{tot}}\!=\!\Phi \!+\! \tilde{\Phi}$ before the confinement is removed and the ÒelectricÓ field is ramped up; see Appendix \ref{sect:app} for details on the numerical method. 

As can be seen in Fig.~\ref{Fig_2}(a), we find excellent agreement between the c.m. trajectories from numerical simulations and semiclassical analytics [Eq.~\eqref{eq:taylor_result_2D}], as shown here by dark-blue dots and light-blue solid lines respectively. We note that there is a clear quantitative deviation between these trajectories and the semiclassical result when the perturbing flux is neglected, as indicated by the solid red line. In Figure~\ref{Fig_2}(b), we also verify that a differential measurement of the c.m. trajectories under positive and negative perturbing magnetic flux $x_\text{c.m.}(B_{xy}; t)\!-\!x_\text{c.m.}(-B_{xy}; t)$ could be used to extract an approximate $ \left(  \nu^{x y }_1 \right)_{\rm{est}}^2$, as predicted by Eq.~\eqref{eq:taylor_result_2D}. We note that for these parameters the differential drift would be of the order of a lattice spacing over typical experimental times around $50 \hbar/J$. As the c.m. velocity scales $\propto q^2 B_{xy}$, from the second term of Eq.~\eqref{eq:taylor_result_2D}, this effect would of course be larger if the perturbing magnetic flux is increased and/or the unperturbed flux is reduced.

\section{4D Quantum Hall Responses and Magnetic Perturbations}  \label{sec:4D}

As a further example, we turn to a 4D quantum Hall system, for which magnetic perturbations enter both into the current and the particle density. A four-dimensional system is also richer than a two-dimensional one as the Berry curvature (\ref{eq:berry2form}) can now have up to six independent components $\Omega^{\mu \nu}$: one for each possible 2D plane. To simplify the following discussion, we therefore first introduce a minimal topological lattice model, noting that to have a non-zero 2CN (\ref{eq:second_chern}) it is sufficient to have only two non-zero Berry curvature components provided they are in ``disconnected" planes. For instance, one can take the non-zero components to be $\Omega^{x z}\!=\!-\Omega^{z x}$ and $\Omega^{y w}\!=\!-\Omega^{w y}$, without loss of generality.

\subsection{Minimal 4D Topological Model} \label{sec:minimal4D}

As we have previously proposed in Refs.~\cite{4Datoms:2015} and~\cite{4Dphotons:2015}, a minimal 4D lattice model that has energy bands with non-zero 2CNs consists of two copies of the Harper-Hofstadter model \cite{Hofstadter} defined in the disconnected planes $x\!-\!z$ and $y\!-\!w$. The corresponding tight-binding Hamiltonian is 
\begin{align}
\hat H=-J &\sum_{\bs r} c^{\dagger}_{\bs r + a\bs e_x}c_{\bs r}+c^{\dagger}_{\bs r + a\bs e_y}c_{\bs r} \label{4D_ham} \\
& + e^{i 2 \pi \Phi_1 x/a} c^{\dagger}_{\bs r + a\bs e_z}c_{\bs r}+ e^{i 2 \pi \Phi_2 y/a} c^{\dagger}_{\bs r + a\bs e_w}c_{\bs r} + \text{h.c.} \notag ,
\end{align}
where $c^{\dagger}_{\bs r}$ creates a fermion at lattice site $\bs r\!=\!(x,y,z,w)$, and the $x\!-\!z$ and $y\!-\!w$ planes are penetrated by uniform magnetic fluxes $\Phi_{1,2}$, respectively. This model could be realised experimentally by exploiting the concept of  synthetic dimensions \cite{Celi:2014,4Datoms:2015, 4Dphotons:2015}, already introduced in Section \ref{section:harper}. By combining a synthetic dimension with either a 3D optical lattice of ultracold atoms~\cite{4Datoms:2015} or with a 3D coupled-cavity array for photons~\cite{4Dphotons:2015}, experiments would be able to build and explore the physics of an effective 4D lattice. Standard techniques, such as those outlined in Section~\ref{sec:2D}, could then be used to generate the necessary complex Peierls' phase factors for the $x\!-\!z$ and $y\!-\!w$ planes. 

The energy spectrum of this 4D model is given by a Minkowski sum of the energy bands of the two constituent Harper-Hofstadter models~\cite{Kraus:2013,4Datoms:2015}, i.e., 
\begin{align}
\mathcal{E}({\bs k})\!=\!\{ E_1 +E_2 \vert E_1\in \mathcal{E}_{xz}(k_x, k_z), E_2 \in \mathcal{E}_{yw}(k_y, k_w) \} \nonumber.
\end{align}
In particular, we will focus on the lowest 4D band, which for appropriate choices of the uniform magnetic fluxes $\Phi_{1,2}$ is non-degenerate and well-isolated from the higher bands. In this minimal lattice model, the eigenstates making up the 4D energy bands are also characterised by only two non-zero Berry curvature components 
\be
\Omega^{zx}=\Omega^{zx}(k_x, k_z) \ne 0 , \quad \Omega^{yw}=\Omega^{yw}(k_y, k_w)  \ne 0, \label{eq:non-zero}
\ee 
 that only depend on the components of momentum along the considered plane.
Consequently, the integral for the 2CN of the lowest band (\ref{eq:second_chern}) can be performed to find $\nu_2\!=\!\nu_1^{zx}\! \times \! \nu_1^{yw}$~\cite{Kraus:2013,4Datoms:2015}, where we have introduced the 1CNs associated with the $x\!-\!z$ and $y\!-\!w$ planes, respectively. For instance, the 1CN characterizing the $x\!-\!z$ plane is defined as
\be
\nu_1^{z x}= \frac{1}{2\pi} \int_{\mathbb{T}^2}  \, \Omega^{z x} (k_{x}, k_{z}) \, \text{d} k_{x} \text{d} k_{z}= - \nu_1^{x z}.
\ee

\subsection{4D Quantum Hall Response}
 
The semiclassical current density (\ref{eq:main}) for a filled lowest (non-degenerate) band in four dimensions is~\cite{4Datoms:2015}: 
\begin{align}
&j^{\mu}=E_{\nu} \frac{1}{(2\pi)^4} \int_{\mathbb{T}^4} \Omega^{\mu \nu} \text{d}^4k + \frac{\nu_2}{4 \pi^2} \varepsilon^{\mu \alpha \beta \nu} E_{\nu} B_{\alpha \beta} , \label{central_transport_equation}
\end{align}
where the second term now explicitly contains the second Chern number of the lowest band (\ref{eq:second_chern}). In the following, we will choose with no loss of generality the perturbing (synthetic) electric field to be along the $y$ direction, $\bs E\!=\!E_y \bs e^y$. However, in 4D, there are various choices for the orientation of the perturbing magnetic field, which will lead to dramatic differences in center-of-mass observables, as we now discuss.

\subsubsection{Perturbing magnetic flux through a 2D plane without Berry curvature}\label{section:case1}

In the simplest case, all non-zero extrinsic perturbing field components $B_{\mu \nu}$ are set in planes for which there is no Berry curvature $\Omega^{\mu \nu}$ from the underlying lattice.  For example, for the minimal model introduced above [Eqs. \eqref{4D_ham}-\eqref{eq:non-zero}], this could be   $B_{z w}\!\ne\!0$, since $\Omega^{zw}\!=\!0$ by construction. For this choice of perturbing magnetic field, the modified density of states reduces to the standard form $D({\bs r}, {\bs k})= \frac{1}{(2 \pi)^4}$, as can be seen from Eq.~\eqref{eq:dos}. Therefore, the density of particles  for a filled lowest band is simply [Eq.~\eqref{eq:densityrealistic}]
\begin{eqnarray}
n&=&  \frac{ A_{\text{MBZ}}^{zx} A_{\text{MBZ}}^{yw}}{(2 \pi)^4}  = \frac{1}{V_{\text{cell}}} ,\label{eq:simple_density}
\end{eqnarray}
where $A_{\text{MBZ}}^{\mu \nu}$ is the area of the magnetic BZ in the $\mu\!-\!\nu$ plane, and where $V_{\text{cell}}$ is the four-dimensional magnetic unit-cell volume. For the lattice introduced in Eq.~\eqref{4D_ham}, with rational fluxes $\Phi_1\!=\!p_1/q_1$ and $\Phi_2\!=\!p_2/q_2$, we have the following expressions
\be
A_{\text{MBZ}}^{zx}\!=\! \frac{(2 \pi)^2}{a^2 q_1} , \, \, A_{\text{MBZ}}^{yw}\!=\! \frac{(2 \pi)^2}{a^2 q_2}, \, \,  V_{\text{cell}}\!=\!q_1 q_2 a^4. 
\ee
Importantly, from Eqs.~\eqref{eq:atomcom} and \eqref{eq:simple_density}, we obtain that the c.m. velocity is  directly proportional to the current density up to a constant factor~\cite{4Datoms:2015},
\be
{\bs v}_\text{c.m.}= \frac{{\bs j}}{n} = {\bs j} V_{\text{cell}}. \label{simple_c_m}
\ee
Here, the current density ${\bs j}$  is explicitly given by [Eq.~\eqref{central_transport_equation}]
\begin{align}
&j^x=  \frac{\nu_2}{4 \pi^2} E_{y} B_{zw}, \label{non_lin_response}\\
&j^{w}=E_{y} \frac{1}{(2\pi)^4} \int_{\mathbb{T}^4} \Omega^{w y} \text{d}^4k = - \frac{\nu_1^{yw} A_{\text{MBZ}}^{ z x}}{(2 \pi)^3} E_{y}, \label{lin_response}\\
&j^y=j^z=0,
\end{align}
where we have used that $\Omega^{wy}\!$ is a function only of the momenta in the $y\!-\!w$ plane in order to perform the integral in Eq.~\eqref{lin_response}. The current response in Eq.~\eqref{non_lin_response} is a non-linear 4D QH response along the $x$ direction, while that in Eq.~\eqref{lin_response} is similar to a 2D QH effect taking place in the $y\!-\!w$ plane. However, unlike the usual 2D QH response (\ref{eq:jx2d}), the current $j^{w}$ is reduced by a factor ${1}/{a^2 q_1}$, as it depends also on the area of the MBZ in the $z\!-\!x$ plane~\cite{4Datoms:2015}.

Combining Eqs.~\eqref{simple_c_m}-\eqref{non_lin_response}, one finds that the center-of-mass displacement along the $x$ direction is directly proportional to the 2CN,
\be
x_\text{c.m.}(t)=\nu_2 \,  (V_{\text{cell}} t/4 \pi^2) E_{y} B_{zw}.\label{c.m._PRL}
\ee
Hence, in this configuration, measurements of center-of-mass observables [Eq.~\eqref{eq:atomcom}\&\eqref{eq:photon}] in atomic or photonics systems can be used to directly extract the 2CN-response, as proposed in Refs.~\cite{4Datoms:2015} and~\cite{4Dphotons:2015}.  The semiclassical predictions in Eqs.~\eqref{non_lin_response}-\eqref{c.m._PRL} have already been validated through numerical simulations in Ref.~\cite{4Datoms:2015}.

\subsubsection{Perturbing magnetic flux through a 2D plane with Berry curvature}\label{section:case2}

In this case, there is only one non-zero extrinsic perturbing field component $B_{\mu \nu}$, and this is in the same plane as a non-zero Berry curvature $\Omega^{\mu \nu}$ from the underlying lattice. For the minimal model proposed above [Eq.~\eqref{eq:non-zero}], this would be the case, for example, when $ B_{zx}\!\ne\!0$ as $\Omega_{zx}\!\ne\!0$. For such a configuration, the density of states is strongly modified [Eq.~\eqref{eq:dos}], and the particle density for a filled band becomes [Eq.~\eqref{eq:densityrealistic}]
\begin{eqnarray}
n&=&   \int_{\mathbb{T}^4} \frac{\text{d}^4k}{(2\pi)^4}  \left( 1 + B_{zx}{\Omega}^{zx} \right)   \nonumber \\
 &=& \frac{ A_{\text{MBZ}}^{zx} A_{\text{MBZ}}^{yw}}{(2 \pi)^4} +  \frac{ A_{\text{MBZ}}^{yw} }{(2\pi)^3}  B_{zx} \nu_1^{zx}  , \notag \\
&=& \frac{ A_{\text{MBZ}}^{yw} }{(2\pi)^2}  \left [ \frac{ A_{\text{MBZ}}^{zx}}{(2 \pi)^2} +  \frac{B_{zx}}{2 \pi} \nu_1^{zx}  \right ]  \label{eq:dens2}
\end{eqnarray}
where we again used that first-order corrections to the Berry curvature vanish upon integration.
Up to the overall factor $A_{\text{MBZ}}^{yw} / (2 \pi)^2$, this is the same modification of the density as found in a 2D system [see Eq.~\eqref{eq:n1}]; this was expected because the Berry curvature component $\Omega^{zx}\!=\!\Omega^{zx}(k_x, k_z)$ only depends on the momenta along the considered $x-z$ 2D plane when performing the integral (see Ref.~\cite{4Datoms:2015}). 

In this case, the current density becomes [Eq. \eqref{central_transport_equation}]
\begin{align}
&j^x=j^y=j^z=0, \notag \\
&j^{w}= - \frac{\nu_1^{yw} A_{\text{MBZ}}^{ z x}}{(2 \pi)^3} E_{y} - \frac{\nu_2}{4 \pi^2} E_{y} B_{zx},\label{response2} 
\end{align}
where now both the 2D-like quantum Hall effect (1CN-response) and the 4D quantum Hall effect (2CN-response) occur along the $w$ direction. These responses could be separated by a differential current measurement, where the sign of the perturbing magnetic field is flipped. For instance, the 2CN can still be extracted from the differential current 
\be
\delta j^w\!=\! j^w(-B_{zx})\!-\!j^w(B_{zx})= \frac{\nu_2}{2 \pi^2} E_{y} B_{zx}. \label{diff_current}
\ee

Surprisingly, if we now turn to the c.m. velocity, we find that the non-zero velocity component $v_\text{c.m.}^w$ simplifies to the expression [Eqs.\eqref{eq:atomcom},\eqref{eq:dens2}-\eqref{response2}]
\begin{align}
&v_\text{c.m.}^w= \frac{j^w}{n} =- \frac{ 2 \pi }{A_{\text{MBZ}}^{yw}} E_{y}  \nu_1^{yw}.\label{vw_anomalous}
\end{align}
Importantly, the c.m. velocity in Eq.~\eqref{vw_anomalous} contains only the linear 1CN-response; all effects from the perturbing magnetic field and 2CN have cancelled out. Hence, in contrast to the current densities in Eqs. \eqref{response2}-\eqref{diff_current}, the c.m. displacement cannot be exploited to extract the 2CN of the populated band. Similar to Examples I and II above, we emphasise these differences by presenting two examples:

 \paragraph*{Example III:} Let us first illustrate Eqs.~\eqref{response2}-\eqref{vw_anomalous} on a simple example: a 4D lattice pierced by a uniform flux $\Phi_{1}\!=\!1/5$ in the $x\!-\!z$ plane and $\Phi_{2}\!=\!1/4$ in the $y\!-\!w$ plane. For this system, the lowest band of the spectrum has a second Chern number $\nu_2\!=\!-1$ and first Chern numbers $\nu^{yw}_1\!=\! -1$ and $\nu^{zx}_1\!=\!1$ in individual planes \cite{4Datoms:2015}, while the magnetic Brillouin zone areas are $A^{yw}_{\text{MBZ}}\!=\!(2\pi/a)^2/4$ and $A^{zx}_{\text{MBZ}}\!=\!(2\pi/a)^2/5$. Let us now assume that there is no perturbing magnetic field $B_{zx}\!=\!0$, in which case the current density from Eq.~\eqref{response2} is $j^{w}\!=\!(1 / 10 \pi a^2) E_{y}$, and the c.m. velocity from Eq.~\eqref{vw_anomalous} is $v_\text{c.m.}^w\!=\! (4 a^2/ \pi) E_y$.

As in Example I in Section~\ref{sec:2D}, we can reinterpret the flux through the $x-z$ plane as a uniform flux $\Phi_1\!=\!1/4$ perturbed by a weak flux $\tilde{\Phi}\!=\!-1/20$, which corresponds to $B_{zx} \!=\! - B_{xz}\!=\!2 \pi \tilde{\Phi}/a^2\!=-\!\pi/10 a^2$. The unperturbed spectrum of this system is still characterized by the properties detailed above, except that now $A^{zx}_{\text{MBZ}}\!=\!(2\pi/a)^2/4$. One then readily verifies that Eq.~\eqref{response2} again leads to $j^{w}\!=\!(1 / 10 \pi a^2) E_{y}$, while Eq.~\eqref{vw_anomalous} leads trivially to the same c.m. velocity as it is independent of any perturbing magnetic field in the $x-z$ plane. 

\paragraph*{Example IV:} 

\begin{figure}[t!]
\includegraphics[width=7.5cm]{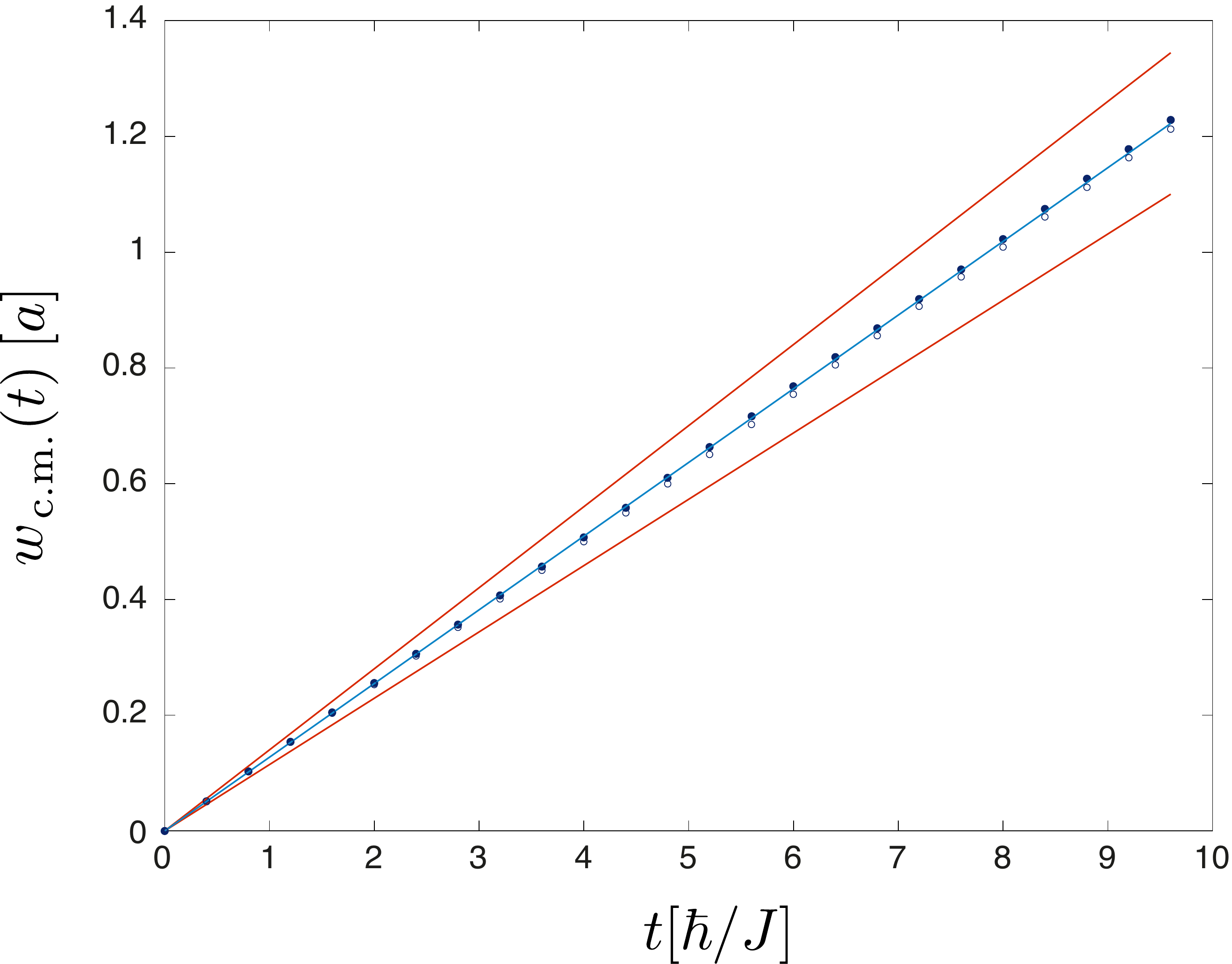}
\vspace{-0.cm} \caption{ Center-of-mass trajectories $w_{\text{c.m.}} (t)$ after ramping up the electric field to $E_y\!=\! 0.2 J/a$, in the presence of strong fluxes $\Phi_{1,2}\!=\!1/4$ in disconnected $x-z$ and $y-w$ planes and a perturbing flux $\tilde{\Phi}\!=\!a^2B_{zx}/2\pi\!=\!\pm 10\% \times  \Phi_1$ in the $x\!-\!z$ plane. The filled dark-blue dots [resp. empty dots] are numerical simulations performed for a small ($4\!\times \!41\! \times \!4\! \times \!41$) 4D lattice with positive [resp. negative] perturbing flux $\tilde{\Phi}$. The light-blue solid curve is the semiclassical analytical result in Eq.~\eqref{vw_anomalous}. The two red dashed curves correspond to the wrong prediction $v_\text{c.m.}^w\!=\! j^w V_{\text{cell}}$ combined with Eq.\eqref{response2}, which corresponds to neglecting the effects of the modified density of states on the particle density \eqref{eq:dens2}. Combining the numerical data together with Eq.~\eqref{vw_anomalous} allows one to extract approximate values for the first Chern number associated with the $y\!-\!w$ plane: we find $(\nu_1^{yw})_{\rm{est}}\!=\!-0.99$ for $\tilde{\Phi}\!>\!0$ and $(\nu_1^{yw})_{\rm{est}}\!=\!-1.00$ for $\tilde{\Phi}\!<\!0$. }\label{Fig_3}
\end{figure}

We have explored the results in Eqs.~\eqref{response2}-\eqref{vw_anomalous} further by performing numerical simulations on a small 4D lattice. Here, the strong fluxes are $\Phi_{1,2}\!=\!1/4$ in disconnected planes \cite{4Datoms:2015}, and we take the perturbing flux to be $\tilde{\Phi}\!=\!a^2B_{zx}/2\pi\!=\!\pm 10\% \times  \Phi_1$ in the $x\!-\!z$ plane. As for the 2D case discussed in Section \ref{sect:2D_c.m.}, an electric field is ramped up to the final value $E_y\!=\! 0.2 J/a$, and the time-evolving particle density is obtained. The resulting center-of-mass trajectories $w_{\text{c.m.}} (t)$ are shown in Fig.~\ref{Fig_3}, which demonstrates good agreement with the prediction in Eq.~\eqref{vw_anomalous}; in particular, one finds that the trajectories show no significant dependence on the perturbing flux. From our numerical data, we use Eq.~\eqref{vw_anomalous} to extract $(\nu_1^{yw})_{\rm{est}}\!=\!-0.99$ for $\tilde{\Phi}\!>\!0$ and $(\nu_1^{yw})_{\rm{est}}\!=\!-1.00$ for $\tilde{\Phi}\!<\!0$, in excellent agreement with the expected 1CN $\nu_1^{yw}\!=\!-1$ of the $y\!-\!w$ plane. The figure also compares the numerical results with the wrong prediction $v_\text{c.m.}^w\!=\! j^w V_{\text{cell}}$, which corresponds to neglecting the effects of the modified density of states on the particle density \eqref{eq:dens2}. We have also numerically verified the expression for the differential current in Eq.~\eqref{diff_current}, which indicates that such measurements could equally be exploited to give an approximate value for the 2CN of the band.

Summarizing the results of Sections \ref{section:case1}-\ref{section:case2}: even though a 4D-QH current response appears for both of the above choices of perturbing magnetic fields [Eqs.~\eqref{non_lin_response} and \eqref{diff_current}], only the first of these is appropriate for extracting the 2CN from center-of-mass observables. This result highlights the importance of evaluating the modified density of states when considering density-dependent observables [Eqs.~\eqref{eq:atomcom}, \eqref{eq:dos}, and \eqref{eq:densityrealistic}].  

\subsubsection{Perturbing magnetic fluxes through two 2D planes with Berry curvature}\label{section:case3}

As a final example, we consider perturbing fields in two planes with non-zero Berry curvature, e.g. $ B_{zx}$ and $ B_{yw}$ for the above model [Eq.~\eqref{eq:non-zero}].  In this case, the particle density for a filled band is strongly modified as [c.f. Eq.~\eqref{eq:densityrealistic}]
\begin{eqnarray}
n&=&   \int_{\mathbb{T}^4} \frac{\text{d}^4k}{(2\pi)^4}  \left( 1 + B_{zx}{\Omega}^{zx} )(1+ B_{yw}{\Omega}^{yw}  \right)   \nonumber \\
 &=& \frac{ A_{\text{MBZ}}^{zx} A_{\text{MBZ}}^{yw}}{(2 \pi)^4} +  \frac{A_{\text{MBZ}}^{yw} B_{zx}   \nu_1^{zx} }{(2\pi)^3} \nonumber \\ && +\frac{A_{\text{MBZ}}^{zx } B_{yw}   \nu_1^{yw}}{(2\pi)^3}  + 
\frac{ B_{zx} B_{y w}}{(2\pi)^2}   \nu_2  
  , \label{eq:dens3}
\end{eqnarray}
 up to second-order in the perturbing fields, where we used that each Berry curvature component $\Omega^{\mu \nu}$ is only a function of momenta in the $\mu\!-\!\nu$ plane. Interestingly, for this configuration of perturbing fields and in contrast to Eq.~\eqref{eq:dens2}, the particle density now explicitly depends on the topological 2CN of the lowest filled band. This suggests an extension of the Streda-Widom formula relating the Hall conductance to density variations with respect to magnetic fields, from 2D~\cite{Widom, Streda, Smrcka} to 4D (see also Supplemental Material of Ref.~\cite{4Datoms:2015}).

 Let us now investigate the current and c.m. responses for this third configuration. Since the perturbing electric field is aligned along the $y$ direction, $\bs E\!=\!E_y \bs e^y$, the component $ B_{yw}$ does not affect the current density, see Eq.~\eqref{central_transport_equation}. Hence, the transport equations obtained in the previous case:
\begin{align}
&j^x=j^y=j^z=0, \notag \\
&j^{w}= - \frac{\nu_1^{yw} A_{\text{MBZ}}^{ z x}}{(2 \pi)^3} E_{y} - \frac{\nu_2}{4 \pi^2} E_{y} B_{zx}, 
\end{align}
are still valid for this configuration. However, using the modified density in Eq.~\eqref{eq:dens3}, the center-of-mass velocity $v_\text{c.m.}^w\!=\!j^w/n$ is now
\begin{eqnarray}
v_\text{c.m.}^w ={\resizebox{.85\hsize}{!}{$\displaystyle{  \frac{ - \nu_1^{yw} A_{\text{MBZ}}^{ z x} E_{y} - 2 \pi \nu_2 E_{y} B_{zx}}{\frac{ A_{\text{MBZ}}^{zx} A_{\text{MBZ}}^{yw}}{2 \pi } +  A_{\text{MBZ}}^{yw} B_{zx}   \nu_1^{zx} +A_{\text{MBZ}}^{zx } B_{yw}   \nu_1^{yw}  + 2 \pi B_{zx} B_{y w}  \nu_2} }$}}, \notag
\end{eqnarray}
which depends on all three topological invariants, $ \nu_1^{zx} $,  $\nu_1^{yw} $ and $\nu_2$. To extract the 4D quantum Hall response and 2CN from such center-of-mass motion would therefore require a multi-step protocol to separate all the different effects. 

\subsubsection{Microscopic Interpretation}

Even though the current density displays a 4D quantum Hall response in all three configurations [Sections \ref{section:case1}-\ref{section:case3}], there can be striking differences in the  topological properties revealed by center-of-mass observables, such as the c.m. velocity of a cloud of ultracold atoms calculated above. We now discuss how this can be understood through a microscopic interpretation of the 4D quantum Hall effect for a filled band of spinless particles. 

From our semiclassical derivation of the current density [Eqs.~\eqref{totalcurrent}-\eqref{eq:main}], we can see that the nonlinear 2CN response arises from combinations of terms in the mean velocity (\ref{eq:meanv}) and in the modified density of states \eqref{eq:dos}. In the first configuration [Sections \ref{section:case1}], as noted above, the density of states \eqref{eq:dos} is not affected by the perturbing magnetic field. Instead, the 2CN response in Eq.~\eqref{non_lin_response} stems entirely from the mean velocity, where it appears from the interplay of the Lorentz force in Eq. \eqref{eq:semik} with the anomalous velocity in Eq. \eqref{eq:semir}, that can itself be interpreted as the analogue of a Lorentz force acting in momentum space. Hence, we refer to this as a ``Lorentz-type" 2CN current response. 

In the second configuration [Sections \ref{section:case2}], on the other hand, the situation is very different: the perturbing magnetic field strongly modifies the density of states [see Eq.~\eqref{eq:dens2}], which then combines with the anomalous velocity in Eq.~\eqref{eq:meanv} to give the 2CN response. In this case, the 4D quantum Hall effect arises from the change in particle density of a band due to a perturbing extrinsic magnetic field, and not from the Lorentz force. We refer to this a ``density-type" 2CN current response. Consequently, when we take into account the particle density for center-of-mass observables, this type of 4D quantum Hall effect vanishes  from the latter [Eq.~\eqref{vw_anomalous}].

In the third configuration [Sections \ref{section:case3}], the 2CN current again arises from the interplay of a change in particle density, induced by an extrinsic field component $B_{zx}$, and the anomalous velocity in Eq.~\eqref{eq:meanv}, i.e. a ``density-type" 2CN response. However, in this case, the second extrinsic magnetic field component $B_{yw}$, while also further changing the particle density, does not couple to the anomalous velocity. When we calculate center-of-mass observables, both changes to the particle density must be taken into account but as only the first leads to a 2CN response, the resulting c.m. observables have a nontrivial dependence on the topological invariants. 

All these configurations could be further combined to yield other more complicated responses, which include both ``Lorentz-type'' and ``density-type'' microscopic mechanisms. Also, we have illustrated our discussion by assuming the Berry curvature 2-form has only two non-zero components, as for the minimal lattice model presented in Section~\ref{sec:minimal4D}. However, the extension to other lattice models would be straightforward.  

\section{Experimental Remarks} \label{sec:experimental}

In this section, we briefly comment on various experimental aspects related to the detection of the quantized responses identified in this work. The platforms that we consider include bosonic and fermionic cold-atomic gases, electrons in solid-state materials and driven-dissipative systems such as coupled photonic cavity arrays~\cite{Hafezi}, where we note that the discussion of the latter can also be extended to coupled electric circuits~\cite{Simon}, or mechanical systems~\cite{Huber}. 

\subsection{Preparation of filled energy bands in the presence of perturbing magnetic fields}

As highlighted in Section~\ref{sec:semiclassical}, we have assumed that the particle density corresponds to that of a system where the lowest band is filled in the presence of any perturbing magnetic fields. In discussing the preparation of such bands, we shall distinguish between three physical scenarios: closed systems in which the particle number is fixed, open systems in which particles can be exchanged with a reservoir, and finally driven-dissipative systems in which an analogue quantum Hall effect can be observed in the long-time non-equilibrium steady state~\cite{Ozawa:2014}.

In closed systems, our assumption of filled bands requires that the initial preparation of the system should be carried out in the presence of all synthetic magnetic fields. 
A standard example of a closed system is an ultracold atomic gas (which is typically not connected to a reservoir). In this case, the assumption of filled bands then means that all magnetic fields should be present during the adiabatic loading of ultracold atoms into the bandstructure.  For ultracold fermions, the energy bands can be filled through fermionic statistics by ensuring that, after loading the atoms, the Fermi energy lies in the middle of an energy band-gap. If instead the perturbing magnetic field were to be ramped up after the atoms were loaded, the Fermi level would end up within an energy band, leading to unquantized anomalous Hall effects~\cite{Xiao2010, Gao}. For thermal Bose gases, an energy band can be uniformly filled when the temperature is large compared with the combined bandwidth of the bands to be populated, as introduced also in Section~\ref{sect:c.m._atom}.  

In open systems, when particles can be exchanged with a reservoir, perturbing magnetic fields can be turned on during the experiment. However, to ensure that the bands are filled in the presence of all magnetic fields, the rate at which these perturbations are ramped up should be slow compared to the time-scale over which particles are exchanged with the reservoirs to ensure equilibration of the particle density. Open systems with reservoirs include solid state materials, in which electrons are exchanged with the connecting leads, but also cold-atom setups involving constriction potentials (see Ref.~\cite{brantut}). Moreover, we point out that harmonically-trapped ultracold fermionic gases potentially present incompressible density plateaus (in a local-density-approximation picture, these correspond to regions with filled bands) surrounded by compressible regions, which can act as an internal reservoir within the cloud~\cite{onur:2008}. Note, however, that such inhomogeneities in the density can also complicate the motion of such a cloud, making it difficult to cleanly observe a quantum Hall response. 

Finally, in driven-dissipative systems, such as coupled photonic cavity arrays, we assume the system is coherently and continuously pumped with sufficient losses to cover the energy bands. Provided that measurements always take place on a sufficiently long time after any variation of the perturbing fields, the photon distribution is in the steady-state and depends only on the current values of these applied perturbations~\cite{Ozawa:2014}. In this case, therefore, perturbing magnetic fields do not need to be ramped up but can be switched on suddenly during the experiment. 

\subsection{Switching-on of the perturbing electric field and the validity of the semiclassical approach}

After the filled energy bands have been prepared, the electric field can be turned on to measure the quantum Hall response. Again we can distinguish between the different physical scenarios introduced above. 

On the one hand, for closed systems and open systems with reservoirs, the electric field should be ramped up adiabatically such that the speed of the ramp is sufficiently slow and that the final value of the fields are sufficiently small compared to the energy band-gap such that diabatic inter-band transitions can be neglected. If these conditions are not fulfilled, then there can be additional contributions to the transport from excited bands, and the current density or c.m. responses will in general no longer be quantised in terms of topological invariants~\cite{Xiao2010, Gao}. For systems with reservoirs, we also require that the time-scale over which the electric field is ramped up is fast compared with the rate with which particles are exchanged, otherwise the equilibration of the system will suppress the quantum Hall response. On the other hand, for driven-dissipative systems, the perturbing electric field can be switched on suddenly, provided that afterwards we wait a sufficiently long time for the steady-state to be reached as discussed above. 

Finally, we emphasise that the semiclassical approach presented in this work can only capture dynamics after any external fields are fully switched on as we have assumed throughout this work that all fields are time-independent. We note that the agreement between the semiclassics and the dynamics after an adiabatic ramping of the electric field is illustrated in Figs.~\ref{Fig_2} \&~\ref{Fig_3}. 

\subsection{Density Measurements and the Streda-Widom formula}

While we have focused in this paper on center-of-mass transport, the dependence of the density on perturbing magnetic fields can itself be used as an experimental tool to measure topological Chern numbers of an energy band, as indicated by the Streda-Widom formula. In ultracold atomic gases, this was first proposed in Ref.~\cite{onur:2008}, where it was demonstrated numerically that the 2D Hall conductance could be extracted by comparing the real space density profile of a cloud for two different values of the total synthetic magnetic field. This protocol could be extended to a 4D system by measuring the density profile of a cloud, for example, for different values of the magnetic fields of the minimal model introduced in Section~\ref{sec:minimal4D}, see Eq.~\eqref{eq:dens3}. As one of the four dimensions is now synthetic, the real-space density imaging should be extended with an optical Stern-Gerlach measurement to also determine the distribution of atoms in the different internal states~\cite{Mancini:2015, Stuhl:2015}.

\section{Conclusions} \label{sec:conclusions}
 
We have discussed how center-of-mass responses can be used to measure Chern numbers and so to directly probe the topology of energy bands. Center-of-mass observables depend not only on the quantum Hall current density but also on the particle density and so can have a more involved dependence on topological invariants than previously considered. In particular, the particle density is itself sensitive to band topology in the presence of extrinsic magnetic perturbations, enriching the quantum Hall physics that may be explored in ultracold atomic gases and photonic systems. 

In the 2D quantum Hall effect, c.m. observables can depend nonlinearly on the first Chern number of a filled energy band, in striking contrast to the linear dependence expected for electrical conductivity measurements. Such effects may already be observable experimentally, as there are inherent uncertainties in the precise value of magnetic flux imposed~\cite{aidelsburger2013,miyake2013,jotzu2014,Aidelsburger:2015, Mancini:2015,Stuhl:2015,Hafezi}. Additionally, new experiments with larger magnetic perturbations could also be engineered straightforwardly to directly probe the physics discussed here. 

Finally, as a further example of these effects, we have considered recent proposals for the realisation of the 4D quantum Hall effect~\cite{4Datoms:2015, 4Dphotons:2015} in realistic experimental systems. As we have seen, the particular configuration of perturbing fields chosen can have a dramatic impact, even leading to a cancellation of all 4D QH effects in center-of-mass observables, despite the persistence of a clear signature in the current density. A clear understanding of the differences between c.m. observables and current density measurements is therefore crucial to the proper design of future experiments. 

\acknowledgments

The authors thank M. Aidelsburger, I. Bloch, J. Dalibard, M. Dalmonte, A. Dauphin, M.
Lohse, S.  Nascimbene, C. Schweizer, and D.-T. Tran for fruitful discussions. H.M.P., T.O. and I.C. are supported by the ERC through the
QGBE grant, by the EU-FET Proactive grant AQuS,
Project No. 640800, and by the Autonomous Province
of Trento, partially through the project ``On silicon chip
quantum optics for quantum computing and secure communications"
(``SiQuro"). H.M.P was also supported by
the EC through the H2020 Marie Sklodowska-Curie Action,
Individual Fellowship Grant No: 656093 ``SynOptic". O.Z.
acknowledges the Swiss National Foundation for financial
support. N.G. is financed by the FRS-FNRS Belgium and by
the BSPO under PAI Project No. P7/18 DYGEST. %

\appendix
\section{Numerical method for the dynamics of incoherent wave-packets}\label{sect:app}

In this Appendix, we present the numerical method used in Sections \ref{sec:2D}-\ref{sec:4D} to simulate the full time dynamics, and hence confirm the semiclassical predictions. 

\begin{figure}[t!]
\includegraphics[width=9cm]{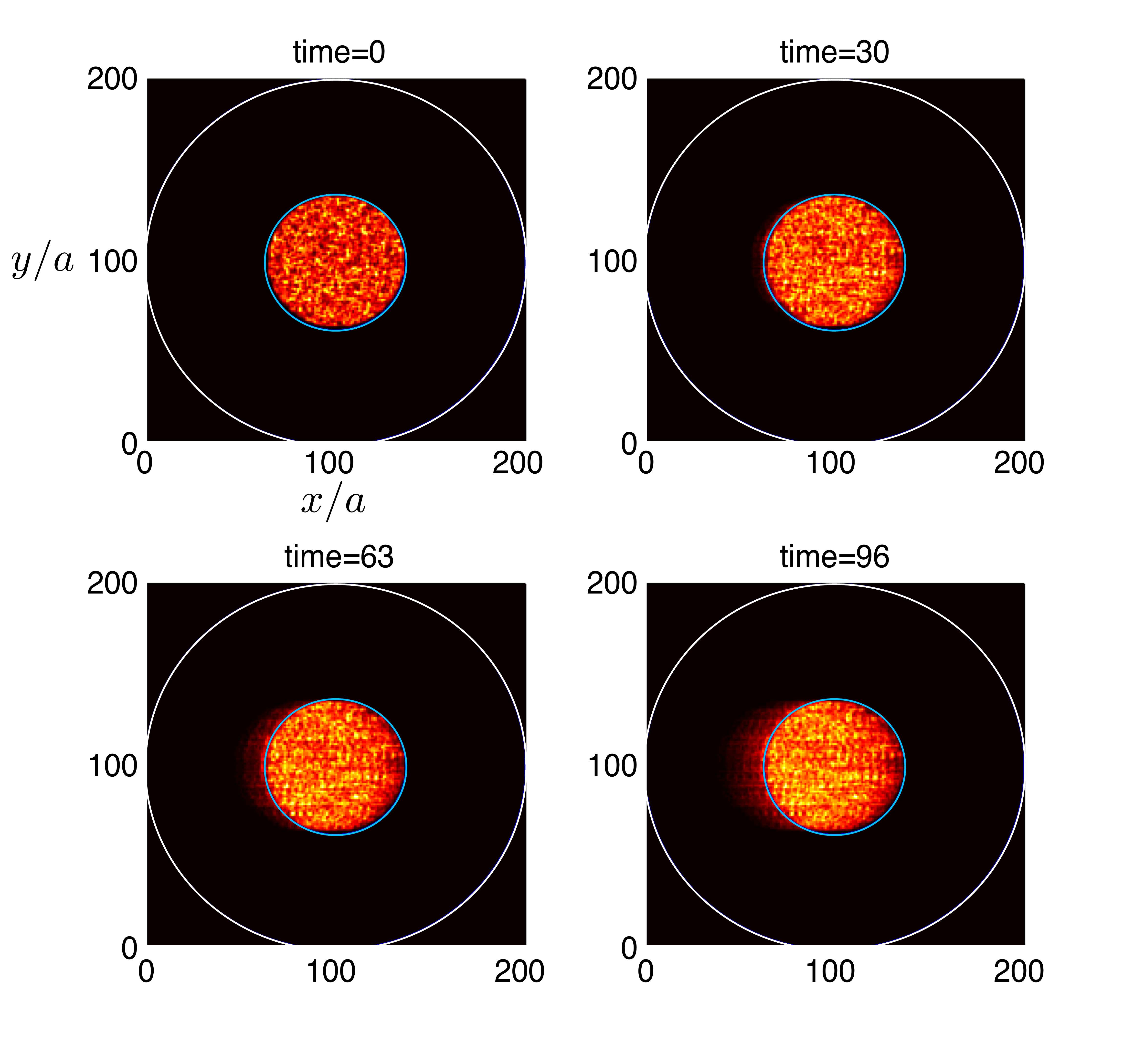}
\vspace{-0.cm} \caption{Time-evolving particle density $n(\bs r,t)$ for the 2D system of Section \ref{sec:2D} subject to a constant electric field $E_y\!=\! 0.2 J/a$ along the $y$ direction. The c.m. drift along the $x$ direction is highlighted by the static circles. Here, the density has been averaged over $N\!=\!8$ realizations. Time is measured in units of $\hbar/J$.
}\label{Fig_app}
\end{figure}

The present method aims to simulate the dynamics of a non-interacting gas that is initially located in a region of space, and which uniformly populates an isolated Bloch band $\mathcal E (\bs k)$ in an incoherent manner. This ``incoherent wave-packet" configuration typically describes an atomic gas whose temperature $T$ is large compared to the bandwidth $W$ of the lowest-energy band $\mathcal E (\bs k)$, but smaller than the band gap, $W\! \ll \! k_{\text{B}}T \! \ll \! \Delta$, as discussed in Section \ref{sect:c.m._atom}. This is different with respect to the numerical method used in Refs.~\cite{Goldman:2013,Dauphin:2013,4Datoms:2015}, which describes a Bloch band completely filled with non-interacting fermions at zero temperature. As already pointed out in Section~\ref{sect:c.m._atom}, these two different band-filling configurations lead to the same semiclassical equations of motion, up to a constant band-filling factor $\rho$ in the current density; we have verified this equivalence numerically by comparing both methods on several examples.

Let us start by building a state from an incoherent superposition of all the Bloch states in a given band, 
\be
\vert \psi_{\text{super}} \rangle=\sum_{\mathcal E_{\lambda} \in \text{band}} \vert \phi_{\lambda} \rangle \exp [i  \theta_{\lambda}] ,
\ee
where $\vert \phi_{\lambda} \rangle$ [resp. $\mathcal E_{\lambda}$] are the single-particle eigenstates [resp. eigenenergies] of the Hamiltonian $\hat H$ in the absence of the perturbing electric field $E_{\mu}$, and where $\theta_{\lambda}$ is a random phase associated with the state $\lambda$. As our lattice models are treated within the tight-binding approximation, we introduce the Wannier basis $\{ \vert j \rangle \}$, which are states localized around the lattice sites $j$. We then define a  closed region in the lattice $\mathcal{S}$, made of a set of lattice sites, and we  project the state $\psi_{\text{super}}$ unto this small region
\be
\vert \psi_{\text{packet}} \rangle= (1/\mathcal{N}) \sum_{j \in \mathcal{S}} \sum_{\mathcal E_{\lambda} \in \text{band}} \vert j \rangle  \langle j \vert \phi_{\lambda} \rangle \exp [i  \theta_{\lambda}],\label{wave packet}
\ee
where $\mathcal{N}$ is a normalization factor. While this projection procedure can also weakly excite particles to higher bands, we have verified that the resulting lowest-band population is between 99\% and 99.9\% in our numerics and so we can safely neglect the contribution from particles in higher bands. The ``wave-packet" in Eq.~\eqref{wave packet} defines the initial state for our simulations. We then act on this state with the time-evolution operator associated with the full Hamiltonian (including the electric field), from which we compute the time-evolving particle density $n (\bs r,t)$, extract the center-of-mass trajectory $r^{\mu}_{\text{c.m.}} (t)$, and determine the corresponding velocity $v^{\mu}_{\text{c.m.}}$. We finally average the results over $N\!\approx\!10$ realizations in which the phases $\theta_{\lambda}$ are randomly generated. The (mean) particle density $n$ and center-of-mass velocities can then be combined to extract the current density, through Eq.~\eqref{eq:atomcom}. Finally, we point out that the band filling factor is simply evaluated as $\rho\!=\!1/N_{\text{states}}$, where $N_{\text{states}}$ is the number of states in the band, i.e. the number of states $\lambda$ included in the sum Eq.~\eqref{wave packet} since the wave packet is normalized. An example of a 2D time-evolving cloud is illustrated in Fig.~\ref{Fig_app}.

\end{document}